%% file: main.tex
\documentclass{article}
\usepackage[utf8]{inputenc}
\usepackage{amsmath}
\usepackage{natbib}
\usepackage{bm}
\usepackage{graphicx}
\usepackage{listings}
\usepackage{longtable}
\usepackage{indentfirst}
\usepackage{longtable}
\usepackage{setspace} 
\usepackage[left=2cm, right=2cm, top=2.00cm, bottom=2.00cm]{geometry}
\title{What events matter for exchange rate volatility?}
\author{Igor Martins \footnote{Örebro University. igor.ferreira-batista-martins@oru.se} and Hedibert Freitas Lopes \footnote{Insper. hedibertfl@insper.edu.br}}
\date{October 2024}

\begin{document}

\maketitle

\section*{Abstract}

This paper expands on stochastic volatility models by proposing a data-driven method to select the macroeconomic events most likely to impact volatility. The paper identifies and quantifies the effects of macroeconomic events across multiple countries on exchange rate volatility using high-frequency currency returns, while accounting for persistent stochastic volatility effects and seasonal components capturing time-of-day patterns. Given the hundreds of macroeconomic announcements and their lags, we rely on sparsity-based methods to select relevant events for the model. We contribute to the exchange rate literature in four ways: First, we identify the macroeconomic events that drive currency volatility, estimate their effects and connect them to macroeconomic fundamentals. Second, we find a link between intraday seasonality, trading volume, and the opening hours of major markets across the globe. We provide a simple labor-based explanation for this observed pattern. Third, we show that including macroeconomic events and seasonal components is crucial for forecasting exchange rate volatility. Fourth, our proposed model yields the lowest volatility and highest Sharpe ratio in portfolio allocations when compared to standard SV and GARCH models.

\vspace{0.5cm}

\noindent \textbf{Key words:} Stochastic Volatility; Macroeconomic Announcements; Sparsity; Seasonality.

\clearpage
\section{Introduction}
Exchange rate volatility has remained a central topic in macroeconomic and finance studies for the past 40 years. Accurate forecasting and understanding of the mechanisms behind it are crucial for policymakers and investors. \cite{blanchard2015can} and \cite{fratzscher2019foreign} highlight the willingness of central bankers to intervene in foreign exchange (FX) markets to smooth exchange rate fluctuations and limit FX volatility, supported by recent theories on welfare gains from such interventions, as in \cite{gabaix2015international}. \cite{bhansali2007volatility} connects carry trading, a strategy based on buying currencies from countries with a high interest rate while selling those from countries with low interest rate, to volatility trading. In his view,  carry trading is profitable during periods of low volatility but performs poorly when volatility spikes. This paper addresses the volatility forecasting challenge and its determinants by modeling the volatility of high-frequency FX returns. We introduce a novel stochastic volatility model that captures the announcement effects of hundreds of macroeconomic variables from multiple countries through spike and slab priors, while also accounting for time-of-day patterns.

Our model captures three main features of intraday returns: volatility persistence, time-of-day effects, and macroeconomic announcements effects. Our methodological contribution comes when modeling announcements. Previous papers, such as \cite{andersen1998answering}, \cite{bauwens2005news}, and \cite{andersen2007real}, select a small number of events based solely on their experience and estimate their effect. This approach may lead to several issues. First, there is a clear possibility of cherry-picking the announcements. Second, by neglecting the inclusion of relevant announcements, estimates of the seasonal or even the persistence component may be affected. Third, it may hinder the identification of relevant macroeconomic channels. Fourth, if the researcher selects irrelevant events, the model is over-parameterized with potential increases in the uncertainty of parameters. Our approach mitigates all of these pitfalls. We start by noting that selecting relevant events solely based on experience is equivalent to assigning probability one to their inclusion and zero otherwise. We relax this approach by allowing the inclusion to be determined not only by the researcher's prior knowledge but also from the data. Specifically, we model the effect of announcements on volatility as coming from a mixture of two distributions: one component is a Dirac delta at zero, reflecting no effect, while the other is a Gaussian distribution with high variance, accommodating a broad range of potential volatility effects. By recovering the probability of an event being included or not, and the effect if the event is included, we access how likely each event is to affect volatility. 

Our model accounts for time-of-day patterns by including dummy variables for each 5-minute window while also accounting for persistence and the effect of macroeconomic announcements. Several papers employ time-of-day effects when modeling intraday FX volatility. \cite{ito2006intraday} observes a U-shaped pattern in both the Japanese Yen and the Euro quoted in US Dollars from 8:00 GMT to 15:00 GMT. \cite{ederington2001intraday}, however, points to the U-shaped pattern on FX markets being due to  macroeconomic announcements on specific days of the week, and after accounting for this feature the U-shaped seasonal effect vanishes. Thus, our proposed model provides a reasonable setting to test the claims of \cite{ito2006intraday}, \cite{ederington2001intraday} as well as other possible patterns. While seasonal effects are common in the intraday literature, they may also play an important role in lower-frequency returns and in other asset classes (e.g., \cite{sorensen2002modeling}).

The final feature of our model is its ability to capture persistence in volatility. Volatility persistence is a common phenomenon in financial markets across both high- and low-frequency settings, and currency returns are no exception (e.g., \cite{andersen1998answering} and \cite{bauwens2005news}). We choose to model volatility persistence using SV models, given their track record of outperforming GARCH in forecasting volatility and enhancing trading strategies for intraday index returns, as shown in \cite{stroud2014bayesian}.

We model 5-minute returns of the Australian Dollar, a currency commonly used as an investment currency in carry trade strategies, as noted by \cite{lustig2011common}. We include all macroeconomic announcements from the US and Australia available in Bloomberg's Economic Calendar, allowing announcement dummies to capture potential impacts on FX volatility. Among hundreds of macroeconomic announcements, the model identifies variables related to the Taylor rule as the only group with over 95\% posterior probability of inclusion. Consistent with \cite{ito2006intraday}, we observe a U-shaped pattern from 8:00 to 15:00 GMT. However, in contrast to \cite{ederington2001intraday} and \cite{ito2006intraday}, we also find an additional U-shaped pattern from 1:30 to 7:00 GMT, resulting in a W-shaped pattern. "Furthermore, our estimated seasonal component is highly correlated with average traded volume, contributing to the literature on volume-volatility connections, as in \cite{abanto2010bayesian}. This new W-shaped pattern may reflect the growth of Asian markets since the early 2000s. Additionally, we link spikes in our estimated seasonal component to major market openings and propose a simple labor economics explanation for the observed pattern. 

We use our estimated model in two applications: volatility forecasting and portfolio allocation. Our proposed model outperforms traditional SV and GARCH models in a realized volatility forecasting exercise. In addition to producing the smallest mean squared errors, Diebold-Mariano tests reject the equal predictability for any common significance level favoring the alternative hypothesis that competitor models are less accurate than our proposal. Furthermore, horse-race regressions show that other models provide little to no additional information once our model is considered. 

We also evaluate our model’s performance in a portfolio allocation problem, where an investor allocates funds between the Swiss Franc and the Australian Dollar. These currencies are commonly used by FX traders in carry strategies as funding and investment currencies, respectively. By combining the volatilities from our model with realized correlations, an investor achieves not only the lowest variance but also the highest Sharpe ratio, outperforming traditional SV and GARCH models. 

The paper is organized as follows. It starts by describing the FX returns, all macroeconomic announcements, our model and estimation approach on Section \ref{Sec:DataModel}. Section \ref{Sec:results} shows our estimates for the macroeconomic announcements, time of the day effects, and volatility persistence. Section \ref{Sec:Applications} presents the volatility forecasting and portfolio allocation applications. Section \ref{Sec:Conclusion} concludes. 

\section{Data and proposed model} \label{Sec:DataModel}

Section \ref{Sec:DataModel} begins by discussing the data used in the empirical applications, including its sources and main characteristics. The section concludes with a description of the proposed model, its priors, and the method for estimating the model by combining the data and priors

\subsection{Data}

The paper models 5-minute returns of the Australian Dollar over a 24-hour period from January 3, 2017, to December 31, 2023, covering 2,554 days. The Australian Dollar, traded 24 hours a day from Sunday at 17:00 Central Time to Friday at 16:00 Central Time, is sourced from FirstRate Data. We use data up to June 29, 2022, for estimation and reserve the remaining observations for out-of-sample analysis. 

We consider 117 macroeconomic events from Australia and the US as potential sources of volatility for the Australian Dollar, as detailed in Appendix A. By including six 5-minute windows after each event, we obtain 702 event-related sources of volatility. Our approach is flexible enough to accommodate events from other countries; however, since the US and Australia provide hundreds of events, far more than any other paper in the literature, we focus exclusively on events from these two countries in this analysis. All timestamps for the macroeconomic announcements are obtained from Bloomberg's economic calendar. We use the lubridate package developed by \cite{grolemund2011dates} to match the timestamps of economic releases with those in the price dataset.

\begin{figure}[h]
\centering
\includegraphics[height=8.5cm,width=0.9\textwidth]{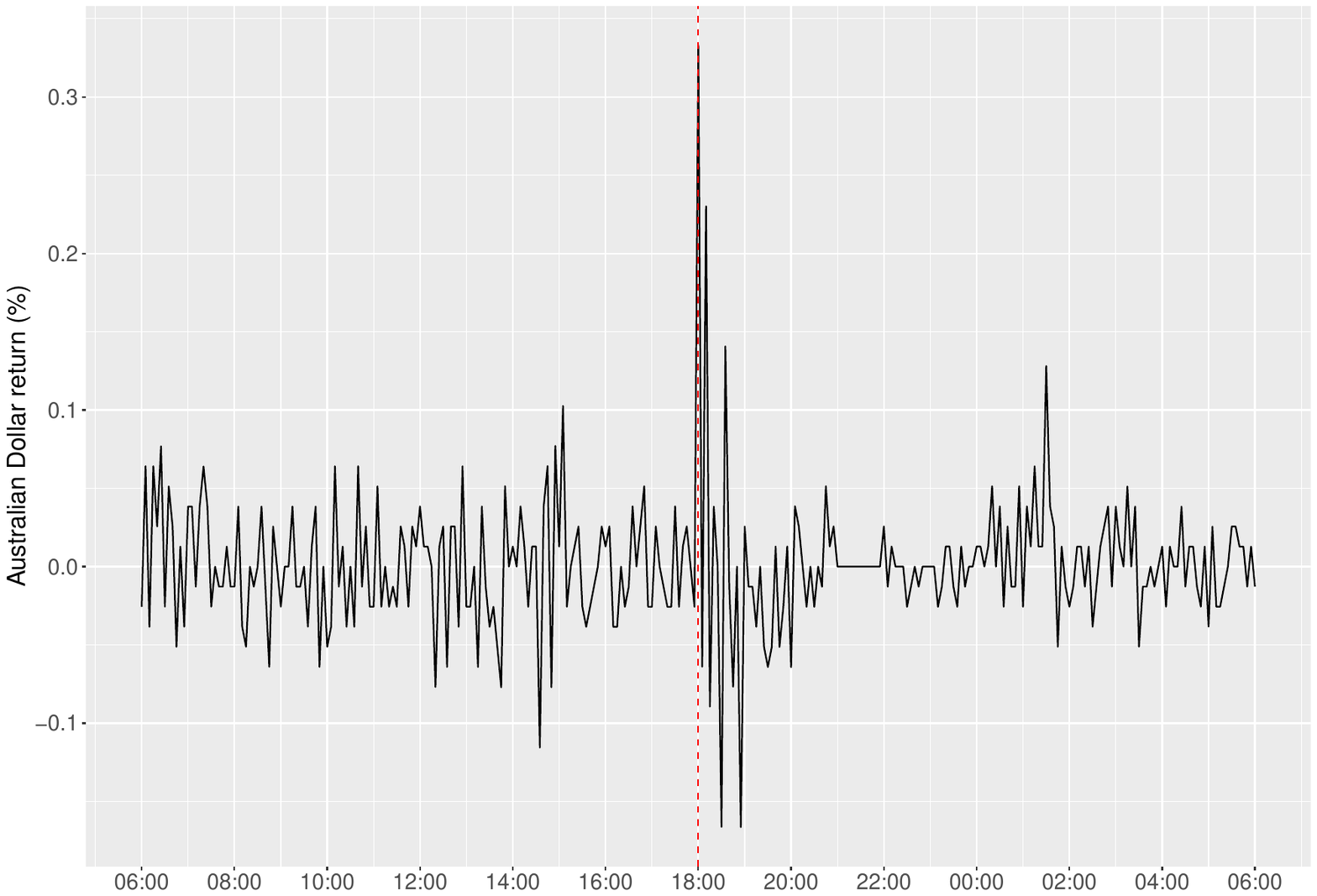}
\caption{
24-hour window of Australian Dollar returns in \% around the FOMC announcement on May 2, 2018. Timestamps are in GMT. The red dashed line indicates the FOMC announcement. Following the announcement, there is a spike in volatility. The spike is large, persists for a few minutes, and then dissipates. Throughout the entire period, a mean of 0 for returns is plausible.}
\label{Fig:FOMC1day}
\end{figure}

Figure \ref{Fig:FOMC1day} illustrates the behavior of the Australian Dollar following an announcement. In particular, it shows the returns of the Australian Dollar during a 24-hour window centered around the FOMC policy announcement on May 02, 2018, marked by a red dashed line. We highlight three plausible features of the data that will be captured in our proposed model. First, the returns have zero mean. Second, volatility spikes after the announcement but quickly dissipates. Third, there is no increase in volatility before the announcement.
Figure \ref{Fig:FOMC1day} is neither an isolated example nor a feature unique to FOMC meetings. Appendix B provides additional examples of the same features around other macroeconomic announcements. 

For our portfolio allocation, we also consider the returns of the Swiss Franc during the same time span, using the same splits for estimation and forecasting as for the Australian Dollar. Similarly, we include macroeconomic announcements for the US and Switzerland when modeling the Swiss Franc. A complete list of the events considered is available in Appendix A. Our choice to model the Australian Dollar, followed by the inclusion of the Swiss Franc in our portfolio application, is due to their critical roles in carry trade strategies. They are commonly used as investment and funding currencies, respectively, as shown in \cite{lustig2011common}.

 \subsection{Model, priors and estimation}
We model 5-min log-returns, $y_t$, allowing for time-varying volatility $v_t$ as shown in Equation (\ref{Eq:ysv}).
\begin{equation}\label{Eq:ysv}
    y_t = v_t \varepsilon_t \text{ with } \varepsilon_{t} \sim N(0,1)
\end{equation}

We follow \cite{stroud2014bayesian} in decomposing the log-variance $h_t$ into the four components presented in Equation (\ref{Eq:logvolspec}): level, stochastic volatility, seasonal and announcement.   
\begin{equation}\label{Eq:logvolspec}
    h_t = log(v_t^2) = \mu_h + x_t + s_t + e_t
\end{equation}

$x_t$ represents the latent persistent stochastic volatility component typical of SV models and is represented by Equation (\ref{Eq:svx})
\begin{equation}\label{Eq:svx}
    x_{t} = \phi x_{t-1} + \sigma_x \eta_{x,t} \text{ with } \eta_{x,t} \sim N(0,1)
\end{equation}

We model the seasonal effect via Equation (\ref{Eq:seasons}). $H_{t}$ is an indicator vector with size 288 capturing all possible 5-minute intervals of the 24 hours of the day. It assumes the value 1 in its k-th component if time t corresponds to the 5 minute window k and 0 otherwise. $\beta_k$ denotes the coefficient associated with the seasonal effect of period k. We restrict the coefficients of the seasonal component to sum zero since we include the level $\mu_h$ as a separate component. 
\begin{equation}\label{Eq:seasons}
    s_t = H_{t}' \beta  \text{ with } 1'\beta = 0 
\end{equation}

We model the announcement effect similarly via Equation (\ref{Eq:anouna}). $I_t$ is an indicator vector whose length equals the number of events being modeled. Each component of the vector takes the value 1 if its corresponding event occurred at time $t$. $\alpha$ is the coefficient that captures the effect of such events. While imposing a decaying effect for lagged events is plausible and possible, our approach treats lags of events as entirely new announcements. Our proposed structure allow events to impact the volatility for different periods. For example, FOMC may impact volatility for a longer period than other announcements. We assume that each event can impact volatility for up to 30 minutes after its release. 
\begin{equation}\label{Eq:anouna}
    e_t = I_{t}' \alpha 
\end{equation}

For a small number of events, the announcement component could be recovered via a Kalman filter without compromising predictive performance. However, given that we consider hundreds of events, a sparsity-inducing approach may improve forecast performance. Additionally, a sparsity-based approach aligns with our belief that some events are irrelevant to prices and volatility. We induce sparsity via spike-and-slab priors for $\alpha$.

The spike and slab prior was introduced by \cite{mitchell1988bayesian}, \cite{george1993variable} and \cite{geweke1996variable} being recently expanded by \cite{ishwaran2005spike}. Its core idea, as presented in Equation (\ref{Eq:spikeslabgeneral}), is to allow each $\alpha_{i}$ from the announcement component to be modeled as having come either from a distribution $p_{spike}(\alpha_{i}|\mathbf{\theta)}$  with mass concentrated around zero or from a distribution $p_{slab}(\alpha_{i}|\mathbf{\theta)}$ with mass covering a long range of values. For example, \cite{george1993variable} consider both $p_{spike}(\cdot)$ and $p_{slab}(\cdot)$ to be Gaussian distributions with different variances.
\begin{equation}\label{Eq:spikeslabgeneral}
    \alpha_{i}|\pi_i \sim  (1 - \pi_i) p_{spike}(\alpha_{i}|\mathbf{\theta_i)} + \pi_i p_{slab}(\alpha_{i}|\mathbf{\theta_i)}
\end{equation}
We consider the spike to be a Dirac delta at zero and the slab to be a Normal with mean 0 and variance $\sigma_{a}^2$. The probability of $\alpha_{i}$ coming from either the spike or the slab is modeled via $\pi_i \sim Bern(\gamma)$. Since $\gamma$ must be between 0 and 1, we assume a beta prior. We use IG prior for variances. For the remainder of the parameters, we assume natural conjugate priors. For both the seasonality coefficients and for the persistence of the AR(1) in Equation (\ref{Eq:svx}), we assume normal priors. Appendix C presents a full description of the priors, hyperpriors, and the MCMC procedure.

We employ a Bayesian approach and use MCMC methods to simulate from the posteriors distribution, i.e., the joint distribution of parameters and latent states conditional on the observed returns $\{ \mu_h, \phi, \sigma^2_x, \{\beta_k\}_{k=1}^{288}, \{\alpha_{i}\}_{i=1}^{I},$ $ \{\pi_i\}_{i=1}^{I}, \gamma, \sigma_a^2, \{x_t\}_{t=1}^{T} \}|y_t$.  Our choice of priors allow for great simplification of the sampling scheme. In summary, we can sample from the posterior of (conditional) linear regression coefficients and variances via Normal and IG distributions. $p(\theta_i|\cdot)$ falls into the Beta - Bernoulli conjugate case leading to a Bernoulli posterior, we can avoid problems due to the Dirac's delta when sampling from $p(\pi_i|\cdot)$ using the approach proposed by \cite{geweke1996variable} and, finally, we can recover $ \{x_t\}_{t=1}^{T}|\cdot$ using \cite{kim1998stochastic} seven Gaussian components approach.   

Our specification for the log-variance in Equation (\ref{Eq:logvolspec}) is consistent with the multiplicative specification given in Equation (\ref{Eq:volspec}). We obtain Equation (\ref{Eq:logvolspec}) by squaring Equation (\ref{Eq:volspec}) and taking logs. 

\begin{equation}\label{Eq:volspec}
    v_t = \sigma X_t S_t E_t 
\end{equation}

While the additive specification is more suitable for estimation,  the multiplicative  specification is easier for interpretation. $\sigma$ represents the volatility level i.e. $v_t$ when $ E_t = S_t = X_t = 1$. $E_t, S_t$ and $X_t$ correspond to the events, seasonal and SV components. Their values are interpreted as by how much you are changing the volatility level. For example, if we have $E_t = 1.1$ with $S_t = X_t = 1$, then events are increasing the volatility level by 10\%. 

\section{Results}\label{Sec:results}

This Section presents the main results in three parts, corresponding to the components in Equation (\ref{Eq:volspec}). We begin by discussing the events most likely to affect volatility and the rationale for their inclusion. Next, we address the seasonal component, showing that the opening of major markets leads to volatility spikes through this component and connecting time-of-day effects to traded volume. Finally, we examine the level and stochastic volatility component. 

\subsection{Macroeconomic announcements}

From all the events considered as possible sources of volatility, only nine are included more than 95\% being five from the USA and four Australia: FOMC rate Decision, US nonfarm payrolls, US CPI, FOMC Meeting Minutes, US retail sales, RBA cash target rate, AU employment change, AU GDP and AU retail sales. The effects on volatility produced by the events up to 30 minutes, i.e. up to six lags, after their release are presented in Figure \ref{Fig:Heatmap}. Both the inclusion and effects of all events are presented in Appendix D.

\begin{figure}[h]
\centering
\includegraphics[height=8.5cm,width=0.9\textwidth]{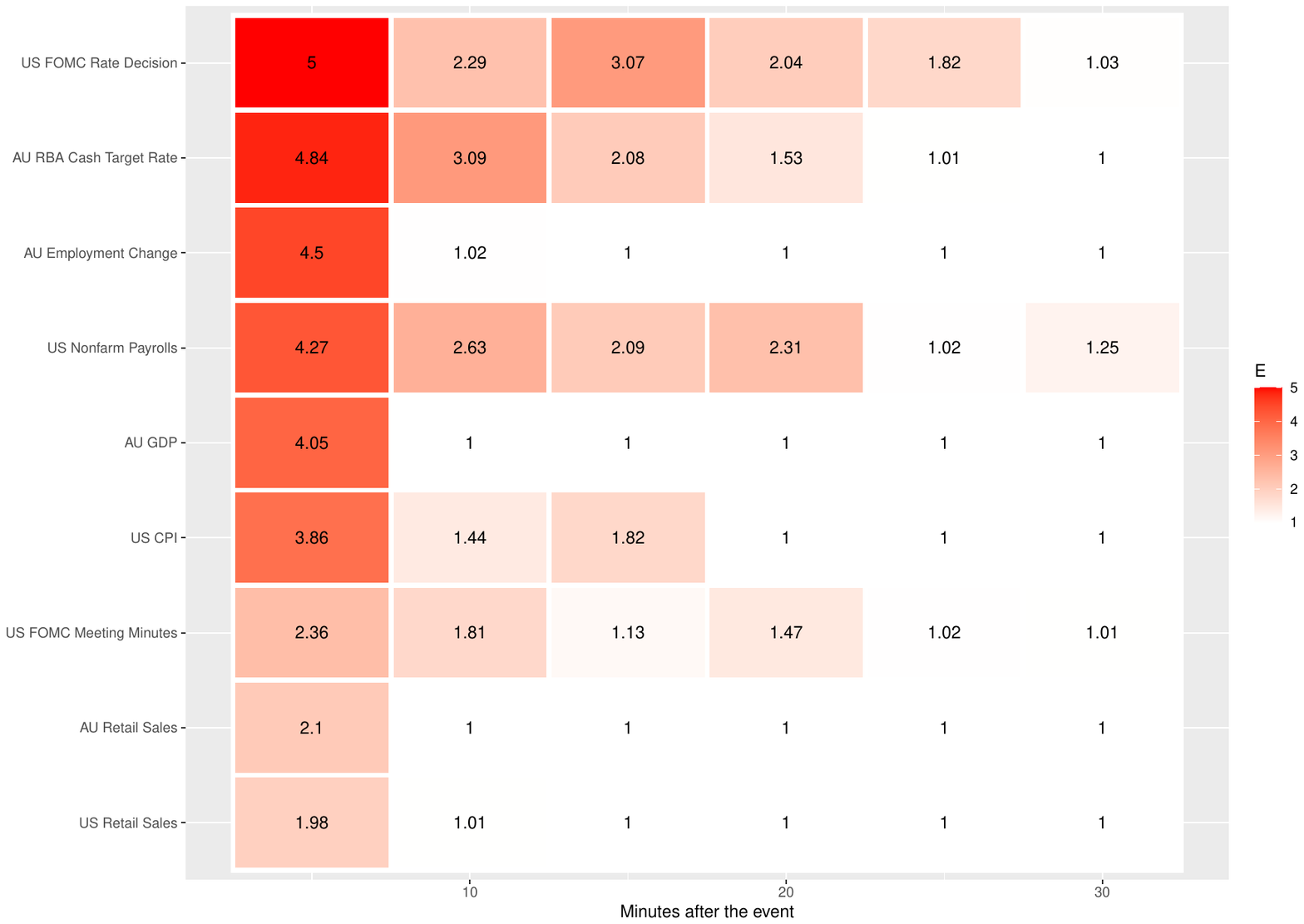}
\caption{Heatmap containing the announcement effect, the posterior mean of $exp\Big(\frac{\alpha}{2}\Big)$, for events with average posterior probability of inclusion higher than 95\% for the first lag. x-axis indicates the number of minute after the announcement occurred capturing the effect on volatility up to 30 minutes after the announcement. Exchange rate volatility is connected to news about macroeconomic fundamentals. All announcements related very likely to be included are related to the Taylor rule.}
\label{Fig:Heatmap}
\end{figure} 

The nine events are informative about variables related to the Taylor rule. \cite{taylor1993discretion} indicates that a central bank sets the interest rates as a function of the difference between current inflation and its target level, and also as a function of the output gap which is the difference between actual and potential output of an economy. Both US and Australia  have their decision about nominal interest rate present in Figure \ref{Fig:Heatmap} represented by the FOMC rate decision and RBA cash target rate, respectively. Furthermore, the FOMC Meeting Minutes 
provides insights on the monetary policy stances of all members of the FOMC committee and is clearly linked to the FOMC rate decision as well. American CPI is used to determine current inflation in the US. GDP and retail sales are informative about the output and therefore for the output gap. Non-farm payroll and employment change are connected to interest rate decisions due to dual mandate of price stability and maximum sustainable employment, as well as their connection with inflation via the NAIRU. Therefore, all of previous events mentioned are connected to macroeconomic fundamentals.

The nine events provide information on variables related to the Taylor rule. \cite{taylor1993discretion} indicates that a central bank sets interest rates based on the difference between current inflation and its target level, as well as the output gap, which is the difference between actual and potential output in an economy. Both the US and Australia have their nominal interest rate decisions represented by the FOMC rate decision and the RBA cash target rate, respectively. Furthermore, the FOMC Meeting Minutes offer insights into the monetary policy stances of FOMC committee members and are closely linked to the FOMC rate decision. The American CPI is used to gauge current inflation in the US. GDP and retail sales provide information on output and, consequently, the output gap. Non-farm payrolls and employment change are relevant to interest rate decisions due to the dual mandate of price stability and maximum sustainable employment, as well as their connection to inflation through NAIRU. Thus, all of the aforementioned events are linked to macroeconomic fundamentals.

We can connect currency returns with our findings based on \cite{campbell1987dollar}. Starting with the definition of log excess currencies returns, \cite{campbell1987dollar} obtain Equation (\ref{Eq:campbellclarida}) connecting nominal exchange rate, $s_t$, via differences of expected nominal interest rates between countries, $E_t \sum_{\tau=0}^{\infty} i_{t +\tau} - i_{t+\tau}^*$, expectations about currency risk premium, $ E_t \sum_{\tau=0}^{\infty} r_{t}^e$, and the expected long-run exchange rate $ E_t \lim_{\tau\to\infty} s_{t + \tau}$. Therefore, exchange rates are connected to interest rate differentials. Since central banks decide interest rates based on Taylor rule variables, news about such variables influence interest rates which in turn influence the exchange rate. 
 
\begin{equation}\label{Eq:campbellclarida}
    s_t = E_t \sum_{\tau=0}^{\infty} (i_{t +\tau} - i_{t+\tau}^*) + E_t  \sum_{\tau=0}^{\infty} r_{t}^e + E_t \lim_{\tau\to\infty} s_{t + \tau}
\end{equation}

 While one may claim that such are events are obvious inclusions, they are often neglected. For example, \cite{marshall2012impact} ignores the possibility of inflation and non-US interest rate decisions affecting volatility. \cite{chen2010news} rules out all non-US events. Additionally, variables that are often included are not supported by our model such as Car sales and Business inventories in the US as included by \cite{bauwens2005news}. Therefore, our model provides a data-driven method to select the events more likely to impact volatility of currencies.

\subsection{Seasonality}
\begin{figure}[h]
\centering
\includegraphics[height=7.5cm,width=0.9\textwidth]{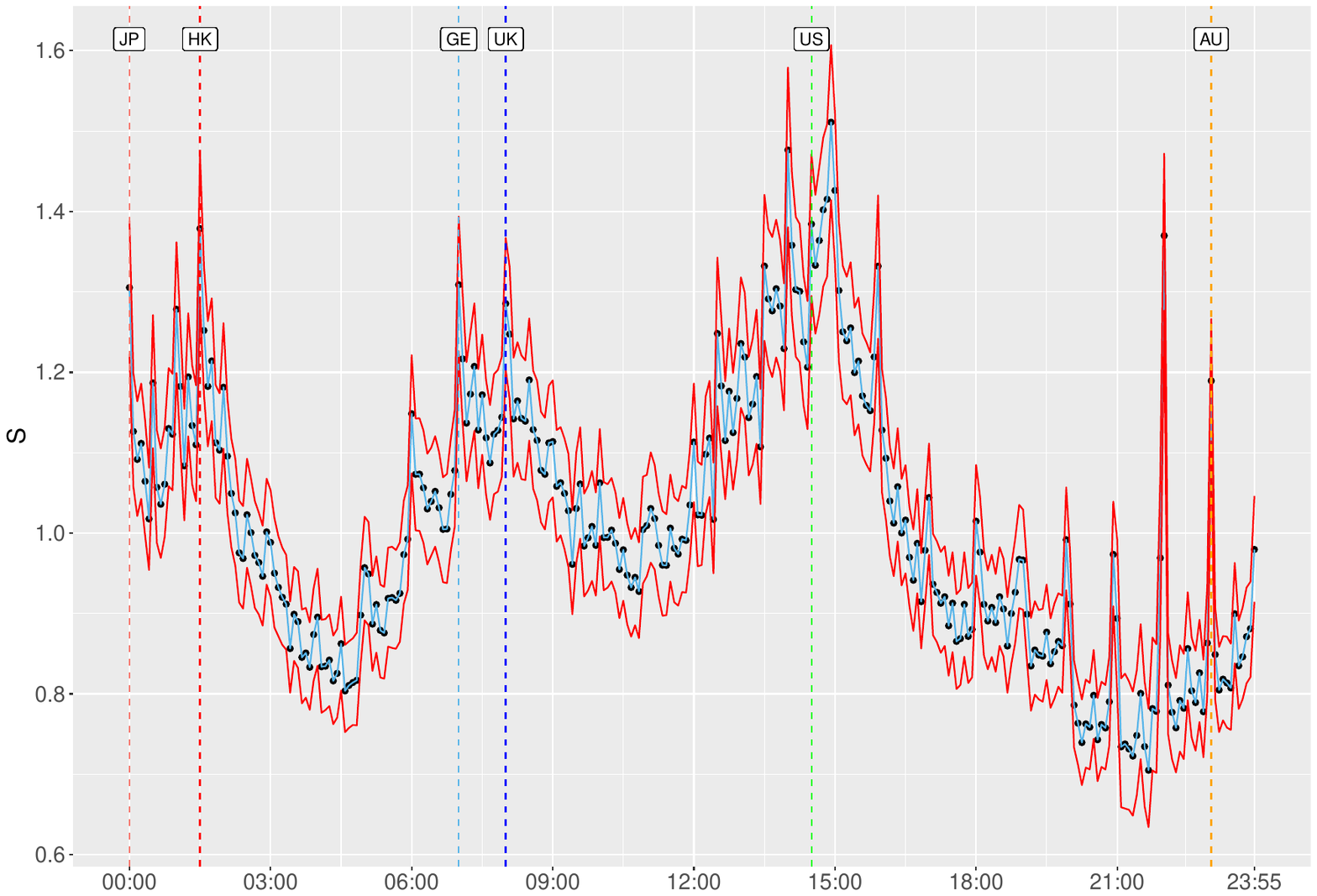}
\caption{Estimated seasonal effect, posterior mean of $S = exp\Big(\frac{s}{2}\Big)$, for the Australian Dollar. An y-axis value of 1.5 indicates that volatility is 50\% higher than its baseline value throughout the day. x-axis represents the time of the day split into 5-minute windows. Black dots represents the posterior mean of each 5 minute interval connected by the solid blue line. Solid red lines indicate a 90\% credible interval for the seasonal effect. Spikes on the seasonality component are linked to opening hours of major exchanges around the globe. For example, Tokyo (00:00 GMT), Hong Kong and Shanghai (01:30 GMT), Frankfurt (07:00 GMT), London (08:00 GMT) and New York (14h30 GMT) show an increase of over 40\% the baseline level while Sidney (23h GMT) has a more modest increase of 4.2\%. The opening hours are represented as dashed vertical lines in pink, red, light and dark blue, green and orange, respectively. The seasonality effect decreases within 90 minutes after the opening for each exchange. There is a notable W pattern peaking on the Chinese, European and American market openings.}
\label{Fig:seasonestimate}
\end{figure}

There are three main results related to the seasonality component. First, we obtain a W-shaped curve for the seasonality effect, the posterior mean of $S = exp\Big(\frac{s}{2}\Big)$, as shown in Figure (\ref{Fig:seasonestimate}). Our estimate shows a distinct U shape starting with the opening hours of the Shanghai and Hong Kong markets (01:30 GMT) and ending with the opening of the German market (07:00 GMT). A second U shape appears, beginning with the opening of the London market (08:00 GMT) and concluding with the opening of the New York Stock Exchange (14:30 GMT). The opening times of these four markets are represented by vertical red, light and dark blue and green dashed lines. This W pattern differs from the typical U-shaped seasonal effects on volatility. For example, \cite{harvey1991volatility}, \cite{hautsch2011econometrics} and \cite{stroud2014bayesian} obtain a single U for the Japanese Yen quoted in dollars, individual assets and for the SP500, respectively, with higher effects during the opening and closing minutes of the US Market. 

Second, the estimated seasonality effects peak when a major exchange opens. While the W-shaped pattern is connected to the opening hours of the Chinese, European, and New York markets, those are not the only markets that affect the estimated seasonal component. For example, at the Tokyo (00:00 GMT) and Australian (23:00 GMT) openings, the seasonal component increases by more than 20\% from the baseline volatility level.

\begin{figure}[h!]
\centering
\includegraphics[height=8cm,width=0.9\textwidth]{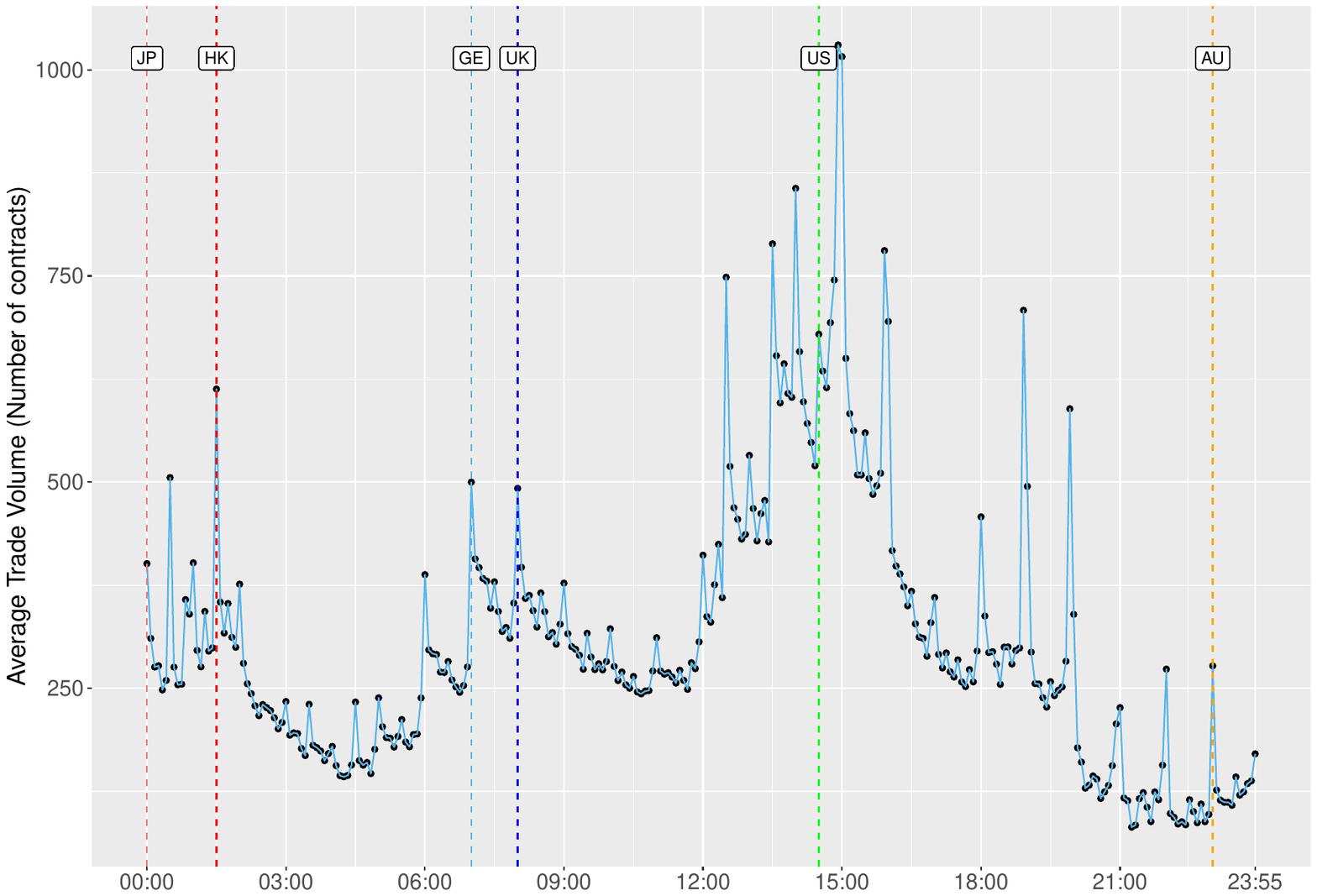}
\caption{ Average traded volume for the Australian Dollar, in number of contracts, for each 5-minute interval within a day. The x-axis shows the time of day split into five-minute windows. Black dots represent the average value connected by a solid blue line. As in the estimated seasonality, we observe spikes at the openings of major exchanges and a W pattern with peaks on the Chinese, European and American market openings. As in Figure(\ref{Fig:seasonestimate}), the opening hours for Tokyo (00:00 GMT), Hong Kong and Shanghai (01:30 GMT), Frankfurt (07:00 GMT), London (08:00 GMT), New York (14h30) and Sydney (23h GMT) are represented as dashed vertical lines in pink, red, light and dark blue, green and orange, respectively. Both the W-shaped pattern and peaks when major markets open are also presented on the average traded volume}
\label{Fig:AvgTradingVolume}
\end{figure}

\begin{figure}[h!]
\centering
\includegraphics[height=8cm,width=0.9\textwidth]{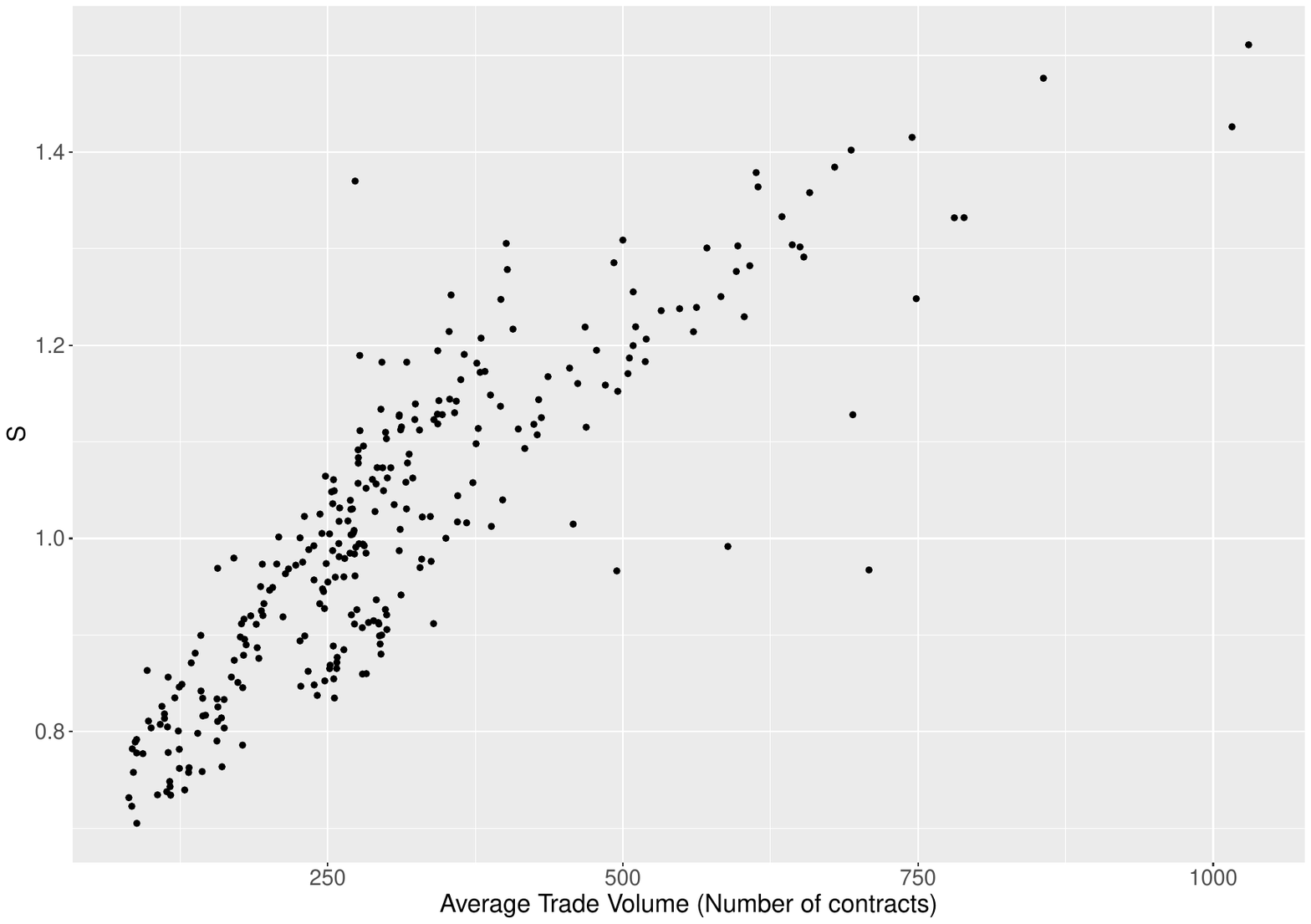}
\caption{ Scatter plot showing the Australian Dollar average traded volume, in number of contracts, for each 5-minute interval within a day on the x-axis and posterior mean of $S = exp\Big(\frac{s}{2}\Big)$ on the y-axis. The plot indicates a clear relationship between estimated seasonality and trading volume presented in Figures (\ref{Fig:seasonestimate}) and (\ref{Fig:AvgTradingVolume}), respectively. A simple linear regression indicates that an increase in the average traded volume of 200 contracts in a 5-minute window is associated with a 55.3\% increase in baseline volatility.}
\label{Fig:Scatterplot}
\end{figure}

Third, while we only use return information in our estimation, our seasonality component is informative about the number of contracts traded. Figure(\ref{Fig:AvgTradingVolume}) plots the average number of contracts traded in each 5-minute window of the day, yielding a shape similar to that presented in Figure(\ref{Fig:seasonestimate}). The similarity between the plots is confirmed by the scatter plot in Figure(\ref{Fig:Scatterplot}) and by a simple linear regression of average traded volume on the posterior mean of the seasonal effects. The regression, with an $R^2$ of 0.88, implies that an increase in the average traded volume of 200 contracts during a 5-minute window is associated with an increase in baseline volatility of 55.3\%.  

The connection between the volatility of returns and trading volume has been noted before by \cite{abanto2010bayesian} in a daily framework. Our findings expand this connection by demonstrating that it is not limited to daily stock returns but is also present across different frequencies and asset classes, with trading volume being largely associated with a specific portion of volatility: the seasonal component.

The spikes in seasonal volatility and trading volume at the opening of a market can partly be attributed to a simple labor-leisure trade-off. Traders begin their working day when a market opens, and higher outputs have been largely associated with initial working hours (e.g., \cite{pencavel2015productivity}). If we consider that a trader knows the amount they should have invested by the end of the day, they may benefit from trading at the beginning of the working day and using part of the remaining hours for leisure purposes.

\subsection{SV component}

\begin{figure}[h]
\centering
\includegraphics[height=8cm,width=0.9\textwidth]{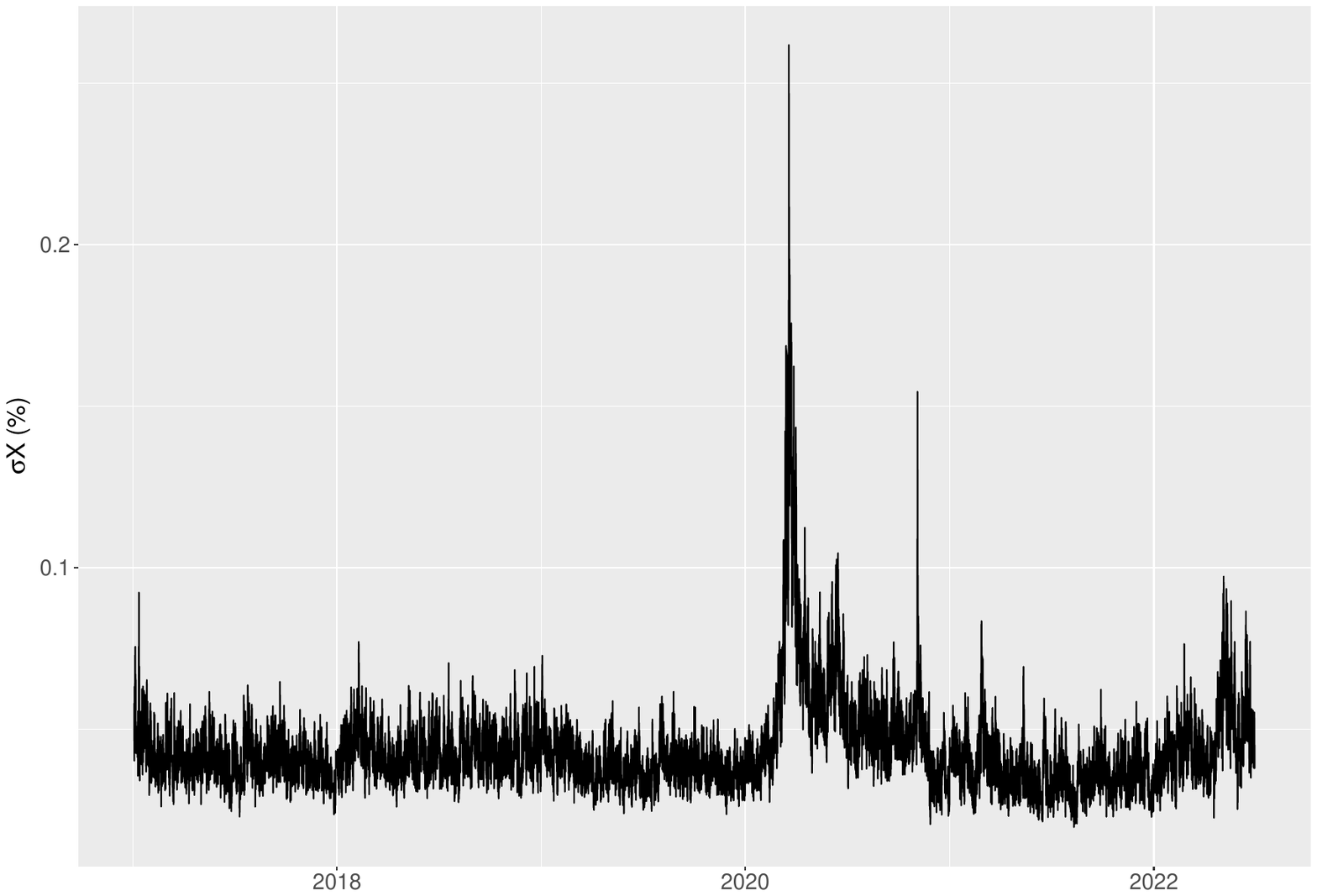}
\caption{Effect of the level and stochastic volatility components represented by the posterior mean of $\sigma X_t = exp\Big(\frac{\mu_h}{2}\Big) exp\Big(\frac{x_t}{2}\Big)$. The rise of volatility on 2020 can be associated with the COVID-19 outbreak.}
\label{Fig:VolPersistence}
\end{figure}

Volatility persistence is a common characteristic in asset returns and is also present in our model. Figure (\ref{Fig:VolPersistence}) plots the combination of the level effect with the SV term, i.e. $\sigma X_t = exp\Big(\frac{\mu_h}{2}\Big) exp\Big(\frac{x_t}{2}\Big)$ when considering the posterior mean for both $\mu_h$ and $x_t$. The estimated posterior mean for the baseline volatility level $\sigma = exp\Big(\frac{\mu_h}{2}\Big)$ is 0.046 (12.5\% on an annualized scale). By interacting with SV component, we observe increases of 5 times the baseline value during the beginning of the COVID19 outbreak in the first quarter of 2020 persisting for months until getting close to its baseline level. 

\section{Forecast and portfolio applications} \label{Sec:Applications}

This section presents two applications: forecasting realized volatility and measuring portfolio risk. It starts by forecasting realized volatility and comparing our proposed model with other commonly used volatility models in the literature. The proposed model yields the smallest mean squared forecasting errors with additional models providing little to no additional forecasting information.  
Next, we consider a second asset, the Swiss Franc, and realized correlations to compare our proposed model with others in a portfolio allocation problem. 

\subsection{Forecasting}

As discussed in \cite{andersen2008realized} and \cite{stroud2014bayesian}, volatility forecasting is crucial for nearly every financial application and is considered the gold standard for evaluating intraday models. We must make a compromise about how to evaluate such forecasts since volatility is unobserved. Our approach is to consider 5-minute realized volatility based on 1-minute returns for the Australian Dollar as our forecasting target.

We compare our forecasting results in two ways: First, following \cite{stroud2014bayesian}, we use horse-race regressions against other models represented by Equation (\ref{Eq:HorseRace}). 

\begin{equation}\label{Eq:HorseRace}
    RV_t = b_0 + b_1 Proposal\_\widehat{Vol_{t|t-1}} + (1 - b_1) Competitor\_\widehat{Vol_{t|t-1}} + \varepsilon_t \text{ with }  b_1 \in [0,1]
\end{equation}

$RV_t$ is the 5-minute realized volatility based on squared 1-minute returns. 
$Proposal_{\widehat{Vol{t|t-1}}}$ is the average volatility forecast by our proposal when parameters are fixed at their posterior mean. $Competitor_{\widehat{Vol_{t|t-1}}}$ is the volatility forecast by other models. If $b_1 = 1$, then competitors provide no additional information for forecasting volatility. 

We consider six competitor models: SSV, SV, AR1-RV, HAR, GARCH, and GJR-GARCH. SSV is the same model as our proposal but without the announcement component, while SV is a traditional stochastic volatility model that does not include seasonality. By comparing our proposal with SSV and SV, we investigate the importance of the announcement effect and the seasonal component. AR1-RV is an AR(1) model for the realized volatility. HAR is a linear model that uses the previous 5-minute, hourly, and daily realized volatility as predictors. GARCH(1,1) represents the traditional Generalized Autoregressive Conditional Heteroskedasticity model, and GJR-GARCH is the Glosten-Jaganathan-Runkle GARCH model, which accounts for potential leverage effects. 

In all cases, our method dominates its competitors by getting $b_1$ close to 1, implying that other methods have almost no additional predictive ability when compared to our proposal, as reported in the first two rows of Table \ref{Tab:HorseR}.

Second, we perform Diebold-Mariano tests of forecasting ability. The null hypothesis is that the two methods have the same forecast accuracy, while the alternative hypothesis is that the competitor is less accurate than our proposal. The third row of Table \ref{Tab:HorseR} reports the p-values of the test, indicating strong evidence against the null hypothesis of equal forecasting accuracy. This leads us to conclude in favor of the alternative hypothesis that competitors are less accurate than our proposal.

\begin{table}[h]
\centering
\begin{tabular}{lcccccc}
  \hline
 & SSV & SV & AR1-RV & HAR & GARCH (1,1) & GJR-GARCH \\ 
  \hline
$b_1$ & 1.00 & 0.95 & 1.00 & 0.99 & 0.99 & 0.99 \\ 
  t-stat & 85.48 & 179.44 & 212.63 & 185.19 & 213.76 & 216.22 \\ 
  \hline
  DM p-value & 0.00 & 0.00 & 0.00 & 0.00 & 0.00 & 0.00 \\ 
   \hline
\end{tabular}
\caption{This table reports two methods of comparing forecast accuracy. In the first two rows, it presents the coefficient an the t-statistic associated with the regression $RV_t = b_0 + b_1 Proposal\_\widehat{Vol_{t|t-1}} + (1 - b_1) Competitor\_\widehat{Vol_{t|t-1}} + \varepsilon_t \text{ with }  b_1 \in [0,1]$. $RV_t$ represents the out-of-sample 5-minute realized volatility based on squared 1-minute returns. $\widehat{Vol_{t|t-1}}$ is the volatility forecasted by a model using information up to t-1. $b_1$ close to 1 indicate that competitor models provide small additional contribution to forecasting realized volatility when compared to our proposal. The last row reports p-values for Diebold-Mariano tests with H0: proposal and competitor have the same forecast accuracy vs H1: competitor is less accurate than the proposal. The near-zero p-values lead us to favor the alternative hypothesis.}
\label{Tab:HorseR}
\end{table}

\subsection{Portfolio application}
 One of the main applications of volatility forecasting is its use as an input for portfolio allocation problems. We consider a global minimum variance portfolio (GMVP) in which a mean-variance investor can choose to take long or short positions in either Australian Dollar or Swiss Franc. The choice of these assets is due to their common use in carry trade strategies, which are based on accruing returns from interest rate differentials between countries. This strategy is largely affected by volatility, as discussed in \cite{bhansali2007volatility}. 
 
 The GMVP problem is a useful tool for evaluating variance-covariance matrices since the weights of each asset don't depend on a model for the average return but only on the volatility of each asset, $vol_i$ and their correlation, $cor_{12}$, as presented in Equation(\ref{Eq:wgmvp}). \cite{campbell2017financial} provides a detailed textbook derivation of Equation (\ref{Eq:wgmvp}). In order to add more realism to the allocation, we also impose a leverage restriction.    
 \begin{equation} \label{Eq:wgmvp}
     w_{1,t}  = \frac{vol_{2,t}^2 - cor_{12,t}}{vol_{1,t}^2 + vol_{2,t}^2 - 2cor_{12,t}} \text{ with } w_{1,t} = 1 - w_{2,t} \text {and }w_{1,t} \in [-1,2]
 \end{equation}
 
Our proposed model and its competitors produce volatility forecasts for individual currencies. However, we must also model the correlation between asset returns. There are multiple reasonable approaches to modeling the dependence structure between asset returns, such as dynamic copulas. In this paper, we take a simple approach and consider the realized correlation over the previous 5 minutes as a reasonable proxy for time-varying correlations.

We compare the performance of our allocation strategy with the same competitor models used in the volatility forecasting subsection. Table \ref{Tab:Portfolio}  presents the results of all allocations. Our model not only yields the smallest volatility but also achieves the highest Sharpe ratio among all models, demonstrating that the selected events have economic implications for investors.

\begin{table}[h]
\centering
\begin{tabular}{rrrrrrrr}
  \hline
 & Proposal & SSV & SV & AR1-RV & HAR & GARCH & GJR-GARCH \\ 
  \hline
Ann. Mean & 8.47 & 7.21 & 2.03 & 2.81 & 5.37 & 6.06 & 5.81 \\ 
Ann. Volatility & 10.50 & 10.49 & 10.80 & 10.76 & 10.64 & 10.66 & 10.65 \\ 
Ann. Sharpe Ratio & 0.81 & 0.69 & 0.19 & 0.26 & 0.50 & 0.57 & 0.55 \\ 
   \hline
\end{tabular}
\caption{Summary statistics of the global minimum variance portfolio for the out-of-sample period, considering our proposal and competitors for an allocation of Australian Dollar and Swiss Franc returns. 'Ann.' indicates annualized values. 'Ann. mean' reports the 5-minute mean multiplied by $\times 252 \times 288$.'Ann. volatility' is the portfolio standard deviation multiplied by  $\times \sqrt{252 \times 288}$. Here, 288 accounts for all 5-minute windows within the 24-hour day, and 252 represents the average number of trading days. The portfolio returns based on our proposed model yield the smallest variance and the highest Sharpe ratio among all models.}
\label{Tab:Portfolio}
\end{table}

\section{Conclusion} \label{Sec:Conclusion}

This paper develops a stochastic volatility model for 5-minute FX returns, accounting for hundreds of macroeconomic events and seasonal components that capture time-of-day effects. Of the possible hundreds of events, only announcements related to Taylor rule variables have more than 95\% inclusion probability for their first lag. We reconcile why news about these macroeconomic fundamentals may affect exchange rates via \cite{campbell1987dollar}. The estimated seasonality effect shows a W-shaped pattern, peaking at the openings of the Chinese, London, and New York markets, while also reflecting volatility due to the openings of other major markets, such as Tokyo and Frankfurt. In addition, our seasonal volatility component is informative about average traded volume. We also demonstrate that the increases in seasonal volatility during major market opening hours correlates with increases in average traded volume. In our forecasting application, our model leads to the most accurate predictions of future realized volatility, with other models providing almost no additional information. Finally, in a portfolio allocation problem, our proposal not only yields the smallest volatility but also the highest Sharpe ratio among all models.







\clearpage
\renewcommand\bibname{References}
\bibliography{biblio}

\clearpage
\section*{Appendix A: Macroeconomic events }

This Appendix presents all the events considered in the empirical applications. Tables \ref{Tab:EvtAu}, \ref{Tab:EvtUs}, and \ref{Tab:EvtCh} show the events for Australia, the US, and Switzerland.

\begin{table}[h]
\begin{tabular}{ll}
\hline
AU AiG Perf of   Construction Index                  & AU Investment   Lending                \\
AU AiG Perf of Mfg Index                             & AU Job Vacancies QoQ                   \\
AU AiG Perf of Services Index                        & AU Manpower Survey                     \\
AU ANZ Job Advertisements MoM                        & AU Melbourne Institute Inflation MoM   \\
AU ANZ Roy Morgan Weekly Consumer Confidence Index   & AU NAB Business Conditions             \\
AU BoP Current Account Balance                       & AU NAB Business Confidence             \\
AU Building Approvals MoM                            & AU Owner-Occupier Loan Value MoM       \\
AU CBA Household Spending YoY                        & AU PPI QoQ                             \\
AU Commodity Index SDR YoY                           & AU Private Capital Expenditure         \\
AU Construction Work Done                            & AU Private Sector Credit MoM           \\
AU Consumer Inflation Expectation                    & AU Private Sector Houses MoM           \\
AU CoreLogic House Px MoM                            & AU RBA 3-Yr Yield Target               \\
AU CPI QoQ                                           & AU RBA Cash Rate Target                \\
AU CPI Trimmed Mean YoY                              & AU RBA FX Transactions Government      \\
AU Credit Card Balances                              & AU RBA Statement on Monetary Policy    \\
AU Employment Change                                 & AU Retail Sales Ex Inflation QoQ       \\
AU Export Price Index QoQ                            & AU Retail Sales  MoM                   \\
AU Exports MoM                                       & AU S\&P Global Australia PMI Composite \\
AU Foreign Reserves                                  & AU S\&P Global Australia PMI Mfg       \\
AU GDP SA QoQ                                        & AU Skilled Vacancies MoM               \\
AU Home Loans MoM                                    & AU Trade Balance                       \\
AU Home Loans Value MoM                              & AU Wage Price Index QoQ                \\
AU House Price Index QoQ                             & AU Westpac Consumer Conf Index         \\
AU Inventories SA QoQ                                & AU Westpac Leading Index MoM           \\
\hline
\end{tabular}
\caption{Table with all Australian events considered in the empirical application}
\label{Tab:EvtAu}
\end{table}

\clearpage
\begin{table}[h]
\begin{tabular}{ll}
\hline
US ADP Employment   Change               & US Kansas City Fed   Manf. Activity              \\
US Advance Goods Trade Balance           & US Kansas City Fed Services Activity             \\
US Business Inventories                  & US Langer Consumer Comfort                       \\
US Challenger Job Cuts YoY               & US Langer Economic Expectations                  \\
US Change in Nonfarm Payrolls            & US Leading Index                                 \\
US Chicago Fed Nat Activity Index        & US MBA Mortgage Applications                     \\
US Conf. Board Consumer Confidence       & US MNI Chicago PMI                               \\
US Construction Spending MoM             & US Monthly Budget Statement                      \\
US Consumer Credit                       & US Mortgage Delinquencies                        \\
US Core PCE Price Index MoM              & US NAHB Housing Market Index                     \\
US CPI MoM                               & US New Home Sales                                \\
US Current Account Balance               & US New York Fed Services Business   Activity     \\
US Dallas Fed Manf. Activity             & US NFIB Small Business Optimism                  \\
US Dallas Fed Services Activity          & US Nonfarm Productivity                          \\
US Durable Goods Orders                  & US NY Fed 1-Yr Inflation Expectations            \\
US Empire Manufacturing                  & US Pending Home Sales MoM                        \\
US Employment Cost Index                 & US Philadelphia Fed Business Outlook             \\
US Existing Home Sales                   & US Philadelphia Fed Non-Manufacturing   Activity \\
US Export Price Index MoM                & US PPI Final Demand MoM                          \\
US Factory Orders                        & US Retail Sales Advance MoM                      \\
US Fed Interest on Reserve Balances Rate & US Richmond Fed Business Conditions              \\
US FHFA House Price Index MoM            & US Richmond Fed Manufact. Index                  \\
US FOMC Meeting Minutes                  & US S\&P CoreLogic CS 20-City MoM SA              \\
US FOMC Rate Decision (Upper Bound)      & US S\&P CoreLogic CS 20-City NSA Index           \\
US GDP Annualized QoQ                    & US S\&P Global US Composite PMI                  \\
US House Price Purchase Index QoQ        & US S\&P Global US Manufacturing PMI              \\
US Household Change in Net Worth         & US S\&P Global US Services PMI                   \\
US Housing Starts                        & US Total Net TIC Flows                           \\
US Import Price Index MoM                & US Trade Balance                                 \\
US Industrial Production MoM             & US Two-Month Payroll Net Revision                \\
US Initial Jobless Claims                & US U. of Mich. Sentiment                         \\
US ISM Manufacturing                     & US U.S. Federal Reserve Releases Beige   Book    \\
US ISM Services Employment               & US Wholesale Inventories MoM                     \\
US ISM Services Index                    & US Wholesale Trade Sales MoM                     \\
US JOLTS Job Openings                    &                                                  \\
\hline
\end{tabular}
\caption{Table with all American events considered in the empirical application}
\label{Tab:EvtUs}
\end{table}

\begin{table}[h]
\begin{tabular}{ll}
\hline
SZ CPI Core YoY                              & SZ Real Estate Index   Family Homes \\
SZ CPI EU Harmonized MoM                     & SZ Retail Sales Real YoY            \\
SZ CPI MoM                                   & SZ SECO Consumer Confidence         \\
SZ Exports Real MoM                          & SZ SNB Policy Rate                  \\
SZ Foreign Currency Reserves                 & SZ SNB Sight Deposit Interest Rate  \\
SZ GDP QoQ                                   & SZ Swiss Watch Exports YoY          \\
SZ Industry \& Construction Output WDA   YoY & SZ Total Sight Deposits CHF         \\
SZ KOF Leading Indicator                     & SZ UBS Real Estate Bubble Index     \\
SZ Money Supply M3 YoY                       & SZ UBS Survey Expectations          \\
SZ PMI Manufacturing                         & SZ Unemployment Rate                \\
SZ Producer \&   Import Prices MoM           &                                     \\
\hline
\end{tabular}

\caption{Table with all Swiss events considered in the empirical application}
\label{Tab:EvtCh}
\end{table}

\clearpage
\section*{Appendix B: Behavior of returns around events}
\begin{figure}[h]
\centering
\includegraphics[height=8.5cm,width=0.9\textwidth]{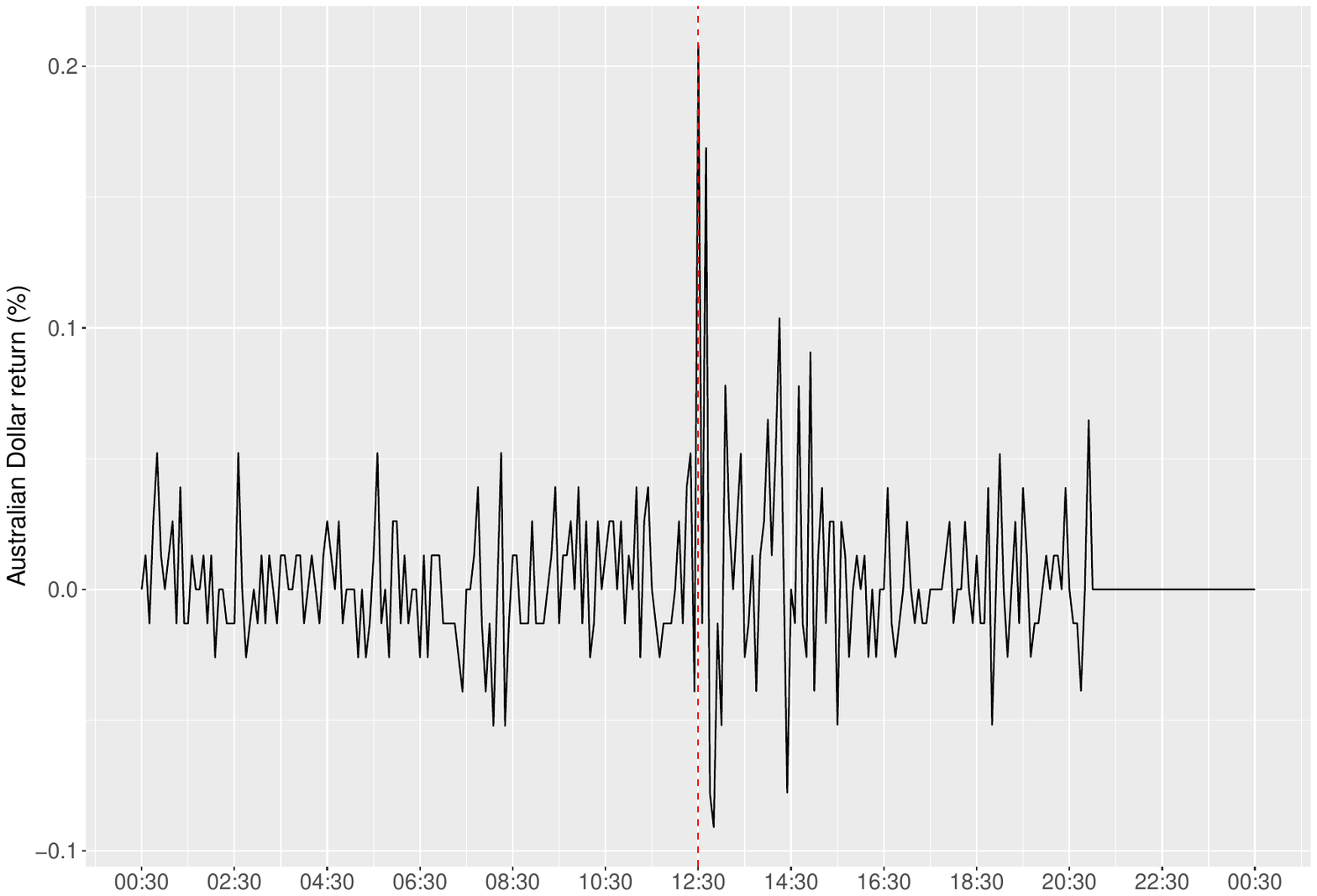}
\caption{
24-hour window of Australian Dollar returns in \% around the Nonfarm Payroll announcement on June 2, 2017. Timestamps are in GMT. The red dashed line indicates the announcement. Following the announcement, there is a spike in volatility. The spike is large, persists for a few minutes, and then dissipates. Throughout the entire period, a mean of 0 for returns is plausible.}
\label{Fig:Payroll1day}
\end{figure}

This Appendix further illustrates the behavior of the Australian Dollar following announcement. It shows that all features highlighted in Section 2 are not restricted to FOMC announcements. Figures \ref{Fig:Payroll1day} and show that following the announcement, there is a spike in volatility. The spike is large, persists for a few minutes, and then dissipates. 

\begin{figure}[h]
\centering
\includegraphics[height=8.5cm,width=0.9\textwidth]{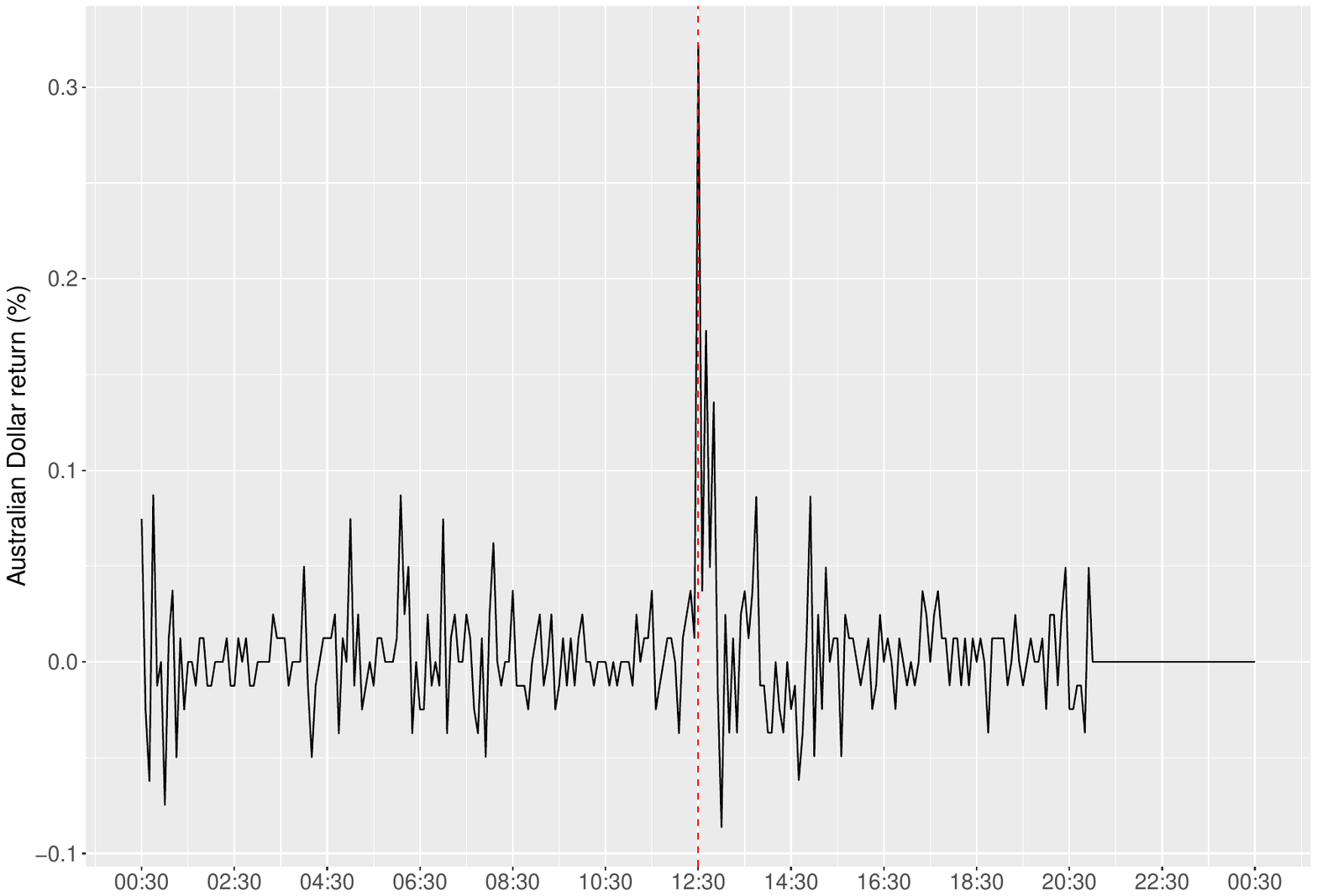}
\caption{
24-hour window of Australian Dollar returns in \% around the CPI announcement on July 14, 2017. Timestamps are in GMT. The red dashed line indicates the announcement. Following the announcement, there is a spike in volatility. The spike is large, persists for a few minutes, and then dissipates. Throughout the entire period, a mean of 0 for returns is plausible.}
\label{Fig:CPI1day}
\end{figure} 

\clearpage
\section*{Appendix C: Priors and MCMC scheme}

This Appendix presents the proposed model, priors and the MCMC scheme employed in this paper. We start by rewriting the main equations of the model described in section 2. We model 5-minute log-returns as having zero mean and time-varying volatility:
$$y_t = v_t \varepsilon_t$$ 

The time varying volatility is composed by the level, SV, seasonal and events components. While we use the additive specification for estimation, we interpret the results based on the multiplicative framework. 
$v_t = \sigma X_t S_t E_t$

$h_t = log(v_t^2) = log(\sigma^2) + log(X_t^2) + log(S_t^2) + log(E_t^2) = \mu_h + x_t + s_t + e_t $

$x_{t} = \phi x_{t-1} + \sigma_x \eta_{x,t}$ 

$s_t = H_{t}' \beta  \text{ with } 1' \beta = 0 $

$e_t = I_{t}' \alpha $

We assume that the events components comes from a Spike and Slab framework. The Dirac Delta represented by $\delta_0$ is the spike and a Gaussian distribution with larger variance acts as the slab. 

$$\alpha_{i}|\pi_i \sim  (1 - \pi_i) \delta_0 + \pi_i N(0, \sigma_{\alpha}^2 \text{ with } \pi_i \sim Be(\gamma) $$  

We must recover $\{  \phi, \sigma^2_x,  \{x_t\}_{t=1}^{T} \}, \mu_h, \{\beta_k\}_{k=1}^{288}, \gamma \{\alpha_{i}\}_{i=1}^{I}, \pi, \sigma^2_{\alpha}|y_t$

We assume standard non-informative priors for all parameters. We attribute Gaussian priors for all coefficients, Inverse Gamma priors for all variances and a beta prior for probabilities. We break down the MCMC methods into the following steps: 

1) $\phi|\cdot$: Gaussian likelihood and Gaussian priors leading to a conjugated Gaussian posterior.   
2) $\sigma^2_x|\cdot$ : Gaussian likelihood and Inverse Gamma prior leading to a conjugated IG posterior. 

3)  $\{x_t\}_{t=1}^{T}\}|.:$ We have a non-linear Gaussian state space problem. We opt to linearize the problem and face a non-Gaussian problem. As in\cite{kim1998stochastic}, we approximate the non -Gaussian problem with a mixture of 7 Gaussian distributions. Using data-augmentation, we can recover the latent states using the Forward Filtering Backward Sampling algorithm of \cite{fruhwirth1994data} and \cite{carter1994gibbs}.

4) $\{\mu_h, \beta\}|.:$ Gaussian likelihood and Gaussian priors leading to a conjugated Gaussian posterior. By using the seasonal dummies as predictor variables, we sample the coefficients of a linear regression without the zero-sum restriction. Than, we recover the mean of those coefficients as $\mu_h$ and the demeaned sample of coefficients as the seasonality $\beta$. 

5) $\gamma|.:$ Binomial model with beta prior leading to a beta posterior.

6) $\sigma^2_{\alpha}|\cdot$ : As in step 2). Gaussian likelihood and IG prior leading to a conjugated IG posterior. 

7) $(\alpha_i, \pi_i|.):$ The update for $\alpha_i$ is straightforward. 0 if $\pi_i=0$ and sample from a conjugate normal if $\pi_i = 1$  due to the normal likelihood and normal distribution of the slab. Updating $\pi_i$ is a bit trickier. The key point of \cite{geweke1996variable} strategy is to integrate over possible values of $\alpha_i$ to avoid problems of the sampler getting stuck on zero due to the infinite mass of the Dirac Delta when $\alpha_i$ is zero.

Denote all parameters with the exception of $\alpha_i$ and $\pi_i$ as $\Xi$
\begin{eqnarray*}
p(\pi_i = 1|\Xi, D) &=& \frac{p(\alpha_i=0, \pi_i = 1|\Xi, D)}{p(\alpha_i=0| \pi_i=1, \Xi, D)} \\
       &=&\frac{p(\Xi, D |\alpha_i=0, \pi_i = 1) p(\alpha_i=0,\pi_i = 1) }{p(\Xi, D) p(\alpha_i=0| \pi_i =1, \Xi, D) } \\
       &=&\frac{p(\Xi, D |\alpha_i=0) p(\alpha_i=0,\pi_i = 1) }{p(\Xi, D) p(\alpha_i=0| \pi_i =1, \Xi, D) } \\
       &\propto& \frac{p(\alpha_i=0, \pi_i = 1) }{p(\alpha_i=0| \pi_i =1, \Xi, D) } \\
       &=& \frac{p(\alpha_i=0|\pi_i = 1)p(\pi_i =1)}{p(\alpha_i=0| \pi_i =1, \Xi, D)} \\
       &=& \frac{\theta \phi(0; 0, \sigma_{\alpha}^2) }{\phi(0; m, v)}
\end{eqnarray*}
where m and v are the  mean and variance of the full conditional posterior distribution for $\alpha_i$ and $\phi(0;a,b)$: Gaussian density at zero  with mean a and variance b. Similarly, for $\pi_i = 0$: 
$$p(\pi_i = 0|\Xi,D) \propto \frac{p(\alpha_i = 0| \pi_i = 0) p(\pi_i =0)}{p(\alpha_i =0 | \pi_i=0, \Xi, D) } = 1 - \theta $$

Therefore, $\pi_i$ can be sampled from the following Bernoulli:

$$Bern\Bigg( \frac{ \frac{\gamma \phi(0; 0, \tau^2) }{\phi(0; m, v)}  }{\frac{\gamma \phi(0; 0, \sigma^2_{\alpha}) }{\phi(0; m, v)} + (1 - \gamma)} \Bigg)$$

\clearpage
\section*{Appendix D: Posterior mean of $\alpha_i$ and $\pi_i$ for the macroeconomic events}

\input{Tables/matrix_betalong.tex}



\input{Tables/matrix_gamma.tex}

\end{document}

%% file: Tables/matrix_betalong.tex
\centering
\begin{longtable}{lrrrrrr}
  \hline
Event Name & 5 Min & 10 Min & 15 Min & 20 Min & 25 Min & 30 Min \\ 
  \hline
AU Commodity Index SDR YoY & 0.00 & 0.00 & 0.00 & 0.00 & 0.00 & 0.00 \\ 
  AU AiG Perf of Services Index & 0.00 & 0.00 & 0.00 & 0.00 & 0.00 & 0.00 \\ 
  AU Trade Balance & 0.13 & 0.00 & 0.00 & 0.00 & 0.00 & 0.00 \\ 
  AU AiG Perf of Construction Index & 0.00 & 0.00 & 0.00 & 0.00 & 0.00 & 0.00 \\ 
  AU Building Approvals MoM & 0.31 & 0.00 & 0.00 & 0.00 & 0.00 & 0.00 \\ 
  AU ANZ Job Advertisements MoM & 0.00 & 0.00 & 0.00 & 0.00 & 0.00 & 0.00 \\ 
  AU Foreign Reserves & 0.00 & 0.00 & 0.00 & 0.00 & 0.00 & 0.00 \\ 
  AU ANZ Roy Morgan Weekly Consumer Confidence Index & 0.00 & 0.00 & 0.00 & 0.00 & 0.00 & 0.00 \\ 
  AU Retail Sales MoM & 1.49 & 0.00 & 0.00 & 0.00 & 0.00 & 0.00 \\ 
  AU Job Vacancies QoQ & 0.00 & 0.00 & 0.00 & 0.00 & 0.00 & 0.00 \\ 
  AU Credit Card Balances & 0.00 & 0.00 & 0.00 & 0.00 & 0.00 & 0.00 \\ 
  AU Melbourne Institute Inflation MoM & 0.00 & 0.00 & 0.00 & 0.00 & 0.00 & 0.00 \\ 
  AU Home Loans MoM & 0.00 & 0.00 & 0.00 & 0.00 & 0.00 & 0.00 \\ 
  AU Investment Lending & 0.00 & 0.00 & 0.00 & 0.00 & 0.00 & 0.00 \\ 
  AU Owner-Occupier Loan Value MoM & 0.00 & 0.00 & 0.00 & 0.00 & 0.00 & 0.00 \\ 
  AU Westpac Consumer Conf Index & 0.00 & 0.00 & 0.00 & 0.00 & 0.00 & 0.00 \\ 
  AU Consumer Inflation Expectation & 0.00 & 0.00 & 0.00 & 0.00 & 0.00 & 0.00 \\ 
  AU Employment Change & 3.01 & 0.04 & 0.00 & 0.00 & 0.00 & 0.00 \\ 
  AU RBA FX Transactions Government & 0.00 & 0.00 & 0.00 & 0.00 & 0.00 & 0.00 \\ 
  AU Westpac Leading Index MoM & 0.00 & 0.00 & 0.00 & 0.00 & 0.00 & 0.00 \\ 
  AU Skilled Vacancies MoM & 0.00 & 0.00 & 0.00 & 0.00 & 0.00 & 0.00 \\ 
  AU CPI YoY & 1.19 & 0.01 & 0.01 & 0.01 & 0.01 & 0.00 \\ 
  AU CPI Trimmed Mean YoY & 2.05 & 0.01 & 0.00 & 0.01 & 0.02 & 0.00 \\ 
  AU PPI QoQ & 0.00 & 0.00 & 0.00 & 0.00 & 0.00 & 0.00 \\ 
  AU Export Price Index QoQ & 0.00 & 0.00 & 0.00 & 0.00 & 0.00 & 0.00 \\ 
  AU NAB Business Conditions & 0.04 & 0.00 & 0.00 & 0.00 & 0.00 & 0.00 \\ 
  AU NAB Business Confidence & 0.02 & 0.00 & 0.00 & 0.00 & 0.00 & 0.00 \\ 
  AU Private Sector Credit MoM & 0.00 & 0.00 & 0.00 & 0.00 & 0.00 & 0.00 \\ 
  AU AiG Perf of Mfg Index & 0.00 & 0.00 & 0.00 & 0.00 & 0.00 & 0.00 \\ 
  AU CoreLogic House Px MoM & 0.00 & 0.00 & 0.00 & 0.00 & 0.00 & 0.00 \\ 
  AU Retail Sales Ex Inflation QoQ & 0.00 & 0.00 & 0.00 & 0.00 & 0.00 & 0.00 \\ 
  AU RBA Cash Rate Target & 3.16 & 2.26 & 1.47 & 0.85 & 0.03 & 0.01 \\ 
  AU Wage Price Index QoQ & 0.03 & 0.00 & 0.00 & 0.00 & 0.00 & 0.00 \\ 
  AU Construction Work Done & 0.00 & 0.00 & 0.00 & 0.00 & 0.00 & 0.00 \\ 
  AU Private Capital Expenditure & 0.10 & 0.00 & 0.00 & 0.00 & 0.00 & 0.00 \\ 
  AU Inventories SA QoQ & 0.00 & 0.00 & 0.00 & 0.01 & 0.00 & 0.00 \\ 
  AU BoP Current Account Balance & 0.00 & 0.00 & 0.00 & 0.00 & 0.00 & 0.00 \\ 
  AU GDP SA QoQ & 2.80 & 0.00 & 0.00 & 0.00 & 0.00 & 0.00 \\ 
  AU Manpower Survey & 0.00 & 0.00 & 0.00 & 0.00 & 0.00 & 0.00 \\ 
  AU House Price Index QoQ & 0.00 & 0.00 & 0.00 & 0.01 & 0.00 & 0.00 \\ 
  AU RBA Statement on Monetary Policy & 0.00 & 0.00 & 0.01 & 0.00 & 0.00 & 0.00 \\ 
  AU S\&P Global Australia PMI Mfg & 0.00 & 0.00 & 0.00 & 0.00 & 0.00 & 0.00 \\ 
  AU S\&P Global Australia PMI Composite & -0.01 & 0.00 & 0.00 & 0.00 & 0.00 & 0.00 \\ 
  AU Home Loans Values MoM & 0.00 & 0.00 & 0.00 & 0.00 & 0.00 & 0.00 \\ 
  AU Private Sector Houses MoM & -0.23 & 0.00 & 0.00 & 0.00 & 0.00 & 0.00 \\ 
  AU Exports MoM & -0.16 & 0.00 & 0.00 & 0.00 & 0.00 & 0.00 \\ 
  AU RBA 3-Yr Yield Target & 0.00 & 0.00 & 0.00 & 0.00 & 0.00 & 0.00 \\ 
  AU CBA Household Spending YoY & 0.00 & 0.00 & 0.00 & -0.01 & 0.00 & -0.01 \\ 
  US S\&P Global US Manufacturing PMI & 0.00 & 0.00 & 0.00 & 0.00 & 0.04 & 0.01 \\ 
  US ISM Manufacturing & 0.24 & 0.10 & 0.03 & 0.00 & 0.00 & 0.00 \\ 
  US Construction Spending MoM & 0.27 & 0.12 & 0.02 & 0.00 & 0.00 & 0.00 \\ 
  US MBA Mortgage Applications & 0.00 & 0.00 & 0.00 & 0.00 & 0.00 & 0.00 \\ 
  US FOMC Meeting Minutes & 1.72 & 1.18 & 0.24 & 0.77 & 0.04 & 0.02 \\ 
  US Challenger Job Cuts YoY & 0.00 & 0.00 & 0.00 & 0.00 & 0.00 & 0.00 \\ 
  US ADP Employment Change & 0.28 & 0.01 & 0.00 & 0.00 & 0.00 & 0.00 \\ 
  US Initial Jobless Claims & 0.22 & 0.05 & 0.00 & 0.00 & 0.00 & 0.00 \\ 
  US S\&P Global US Services PMI & 0.00 & 0.00 & 0.00 & 0.01 & 0.00 & 0.01 \\ 
  US S\&P Global US Composite PMI & 0.00 & 0.00 & 0.00 & -0.01 & 0.00 & -0.01 \\ 
  US Langer Consumer Comfort & 0.00 & 0.00 & 0.00 & 0.00 & 0.00 & 0.00 \\ 
  US ISM Services Index & 0.16 & 0.00 & 0.00 & 0.00 & 0.00 & 0.00 \\ 
  US Trade Balance & 0.00 & 0.00 & 0.00 & 0.00 & 0.00 & 0.00 \\ 
  US Two-Month Payroll Net Revision & -0.02 & -0.01 & 0.34 & 0.00 & 0.05 & 0.62 \\ 
  US Change in Nonfarm Payrolls & 2.91 & 1.94 & 1.47 & 1.68 & 0.04 & 0.44 \\ 
  US Factory Orders & 0.00 & 0.00 & 0.00 & 0.00 & 0.00 & 0.00 \\ 
  US Durable Goods Orders & 0.00 & 0.00 & 0.00 & 0.00 & 0.00 & 0.00 \\ 
  US Consumer Credit & 0.00 & 0.00 & 0.00 & 0.00 & 0.00 & 0.00 \\ 
  US NFIB Small Business Optimism & 0.00 & 0.00 & 0.00 & 0.00 & 0.00 & 0.00 \\ 
  US Wholesale Inventories MoM & 0.00 & 0.00 & 0.00 & 0.00 & 0.00 & 0.00 \\ 
  US Wholesale Trade Sales MoM & 0.00 & 0.00 & 0.00 & 0.00 & 0.00 & 0.00 \\ 
  US JOLTS Job Openings & 0.00 & 0.00 & 0.00 & 0.00 & 0.00 & 0.00 \\ 
  US Import Price Index MoM & 0.00 & 0.00 & 0.00 & 0.00 & 0.00 & 0.00 \\ 
  US Monthly Budget Statement & 0.00 & 0.00 & 0.00 & 0.00 & 0.00 & 0.00 \\ 
  US PPI Final Demand MoM & 0.01 & 0.00 & 0.00 & 0.00 & 0.00 & 0.00 \\ 
  US Retail Sales Advance MoM & 1.37 & 0.01 & 0.00 & 0.00 & 0.00 & 0.00 \\ 
  US Business Inventories & 0.00 & 0.00 & 0.00 & 0.00 & 0.00 & 0.00 \\ 
  US U. of Mich. Sentiment & 0.00 & 0.00 & 0.00 & 0.01 & 0.00 & 0.00 \\ 
  US Empire Manufacturing & 0.00 & 0.00 & 0.00 & 0.00 & 0.00 & 0.00 \\ 
  US CPI MoM & 2.70 & 0.73 & 1.20 & 0.00 & 0.00 & 0.00 \\ 
  US Industrial Production MoM & 0.00 & 0.00 & 0.00 & 0.00 & 0.00 & 0.00 \\ 
  US NAHB Housing Market Index & 0.00 & 0.00 & 0.00 & 0.00 & 0.00 & 0.00 \\ 
  US Total Net TIC Flows & 0.00 & 0.00 & 0.00 & 0.00 & 0.00 & 0.00 \\ 
  US Housing Starts & 0.00 & 0.00 & 0.00 & 0.00 & 0.00 & 0.00 \\ 
  US Philadelphia Fed Business Outlook & 0.00 & 0.01 & 0.00 & 0.00 & 0.00 & 0.00 \\ 
  US Langer Economic Expectations & 0.00 & 0.00 & 0.00 & 0.00 & 0.00 & 0.00 \\ 
  US Existing Home Sales & 0.00 & 0.00 & 0.00 & 0.00 & 0.00 & 0.00 \\ 
  US Richmond Fed Manufact. Index & 0.00 & 0.00 & 0.00 & 0.00 & 0.00 & 0.00 \\ 
  US FHFA House Price Index MoM & 0.00 & 0.00 & 0.00 & 0.00 & 0.00 & 0.00 \\ 
  US Advance Goods Trade Balance & 0.00 & 0.00 & 0.00 & 0.00 & 0.00 & 0.00 \\ 
  US Chicago Fed Nat Activity Index & 0.00 & 0.00 & 0.00 & 0.00 & 0.00 & 0.00 \\ 
  US New Home Sales & 0.00 & 0.00 & 0.00 & 0.00 & 0.00 & 0.00 \\ 
  US Leading Index & 0.00 & 0.00 & 0.00 & 0.00 & 0.00 & 0.00 \\ 
  US Kansas City Fed Manf. Activity & 0.00 & 0.00 & 0.00 & 0.00 & 0.00 & 0.00 \\ 
  US GDP Annualized QoQ & 0.01 & 0.00 & 0.00 & 0.00 & 0.00 & 0.00 \\ 
  US Core PCE Price Index MoM & 0.00 & 0.00 & 0.00 & 0.00 & 0.00 & 0.00 \\ 
  US Pending Home Sales MoM & 0.00 & 0.00 & 0.00 & 0.00 & 0.00 & 0.00 \\ 
  US Dallas Fed Manf. Activity & 0.00 & 0.00 & 0.00 & 0.00 & 0.00 & 0.00 \\ 
  US Employment Cost Index & 0.01 & 0.01 & 0.00 & 0.00 & 0.00 & 0.00 \\ 
  US S\&P CoreLogic CS 20-City MoM SA & 0.00 & 0.00 & 0.00 & 0.00 & 0.00 & 0.00 \\ 
  US S\&P CoreLogic CS 20-City NSA Index & 0.00 & 0.00 & 0.00 & 0.00 & 0.00 & 0.00 \\ 
  US MNI Chicago PMI & 0.00 & 0.00 & 0.00 & 0.00 & 0.00 & 0.00 \\ 
  US Conf. Board Consumer Confidence & 0.00 & 0.00 & 0.00 & 0.00 & 0.00 & 0.00 \\ 
  US FOMC Rate Decision (Upper Bound) & 3.22 & 1.66 & 2.25 & 1.43 & 1.20 & 0.06 \\ 
  US Nonfarm Productivity & 0.00 & 0.00 & 0.00 & 0.00 & 0.00 & 0.00 \\ 
  US Mortgage Delinquencies & 0.00 & 0.00 & 0.00 & 0.00 & 0.00 & 0.00 \\ 
  US House Price Purchase Index QoQ & 0.00 & 0.00 & 0.00 & 0.00 & 0.00 & 0.00 \\ 
  US U.S. Federal Reserve Releases Beige Book & 0.00 & 0.00 & 0.00 & 0.00 & 0.00 & 0.00 \\ 
  US Household Change in Net Worth & 0.00 & 0.00 & 0.00 & 0.00 & 0.00 & 0.00 \\ 
  US Current Account Balance & 0.00 & 0.00 & -0.01 & 0.00 & 0.00 & 0.00 \\ 
  US Export Price Index MoM & 0.00 & 0.00 & 0.00 & 0.00 & 0.00 & 0.00 \\ 
  US Fed Interest on Reserve Balances Rate & 0.00 & 0.00 & -0.02 & 0.09 & 0.09 & 0.07 \\ 
  US ISM Services Employment & 0.00 & -0.02 & 0.06 & -0.04 & -0.02 & 0.05 \\ 
  US New York Fed Services Business Activity & -0.03 & -0.01 & 0.04 & 0.01 & -0.04 & -0.02 \\ 
  US Philadelphia Fed Non-Manufacturing Activity & 0.03 & -0.01 & 0.05 & 0.02 & 0.02 & -0.01 \\ 
  US Richmond Fed Business Conditions & 0.01 & 0.04 & -0.03 & -0.02 & -0.03 & 0.03 \\ 
  US Kansas City Fed Services Activity & 0.03 & -0.03 & 0.00 & 0.01 & -0.04 & 0.04 \\ 
  US Dallas Fed Services Activity & 0.06 & 0.02 & -0.05 & -0.04 & -0.05 & -0.02 \\ 
  US NY Fed 1-Yr Inflation Expectations & -0.02 & 0.01 & -0.06 & -0.02 & -0.02 & 0.00 \\ 
   \hline
   \caption{Posterior mean of $\alpha$ for all events}
\end{longtable}

%% file: Tables/matrix_gamma.tex

\centering
\begin{longtable}{lrrrrrr}
  \hline
Event Name & 5 Min & 10 Min & 15 Min & 20 Min & 25 Min & 30 Min \\ 
  \hline
AU Commodity Index SDR YoY & 0.00 & 0.00 & 0.00 & 0.00 & 0.00 & 0.00 \\ 
  AU AiG Perf of Services Index & 0.00 & 0.00 & 0.00 & 0.00 & 0.00 & 0.00 \\ 
  AU Trade Balance & 0.09 & 0.00 & 0.00 & 0.00 & 0.00 & 0.00 \\ 
  AU AiG Perf of Construction Index & 0.00 & 0.00 & 0.00 & 0.00 & 0.00 & 0.00 \\ 
  AU Building Approvals MoM & 0.24 & 0.00 & 0.00 & 0.00 & 0.00 & 0.00 \\ 
  AU ANZ Job Advertisements MoM & 0.00 & 0.00 & 0.00 & 0.00 & 0.00 & 0.00 \\ 
  AU Foreign Reserves & 0.00 & 0.00 & 0.00 & 0.00 & 0.00 & 0.00 \\ 
  AU ANZ Roy Morgan Weekly Consumer Confidence Index & 0.00 & 0.00 & 0.00 & 0.00 & 0.00 & 0.00 \\ 
  AU Retail Sales MoM & 1.00 & 0.00 & 0.00 & 0.00 & 0.00 & 0.00 \\ 
  AU Job Vacancies QoQ & 0.00 & 0.00 & 0.00 & 0.00 & 0.00 & 0.00 \\ 
  AU Credit Card Balances & 0.00 & 0.00 & 0.00 & 0.00 & 0.00 & 0.00 \\ 
  AU Melbourne Institute Inflation MoM & 0.00 & 0.00 & 0.00 & 0.00 & 0.00 & 0.00 \\ 
  AU Home Loans MoM & 0.00 & 0.00 & 0.00 & 0.00 & 0.00 & 0.00 \\ 
  AU Investment Lending & 0.00 & 0.00 & 0.00 & 0.00 & 0.00 & 0.00 \\ 
  AU Owner-Occupier Loan Value MoM & 0.00 & 0.00 & 0.00 & 0.00 & 0.00 & 0.00 \\ 
  AU Westpac Consumer Conf Index & 0.01 & 0.00 & 0.00 & 0.00 & 0.00 & 0.00 \\ 
  AU Consumer Inflation Expectation & 0.00 & 0.00 & 0.00 & 0.00 & 0.00 & 0.00 \\ 
  AU Employment Change & 1.00 & 0.05 & 0.01 & 0.00 & 0.00 & 0.00 \\ 
  AU RBA FX Transactions Government & 0.00 & 0.01 & 0.00 & 0.00 & 0.00 & 0.01 \\ 
  AU Westpac Leading Index MoM & 0.00 & 0.00 & 0.00 & 0.00 & 0.00 & 0.00 \\ 
  AU Skilled Vacancies MoM & 0.00 & 0.00 & 0.00 & 0.00 & 0.00 & 0.00 \\ 
  AU CPI YoY & 0.37 & 0.01 & 0.01 & 0.01 & 0.01 & 0.00 \\ 
  AU CPI Trimmed Mean YoY & 0.63 & 0.01 & 0.01 & 0.01 & 0.01 & 0.00 \\ 
  AU PPI QoQ & 0.00 & 0.00 & 0.00 & 0.00 & 0.00 & 0.00 \\ 
  AU Export Price Index QoQ & 0.00 & 0.00 & 0.00 & 0.00 & 0.00 & 0.00 \\ 
  AU NAB Business Conditions & 0.05 & 0.00 & 0.00 & 0.00 & 0.00 & 0.00 \\ 
  AU NAB Business Confidence & 0.03 & 0.00 & 0.00 & 0.00 & 0.00 & 0.00 \\ 
  AU Private Sector Credit MoM & 0.00 & 0.00 & 0.00 & 0.00 & 0.00 & 0.00 \\ 
  AU AiG Perf of Mfg Index & 0.00 & 0.00 & 0.00 & 0.00 & 0.00 & 0.00 \\ 
  AU CoreLogic House Px MoM & 0.00 & 0.00 & 0.00 & 0.00 & 0.00 & 0.00 \\ 
  AU Retail Sales Ex Inflation QoQ & 0.00 & 0.00 & 0.00 & 0.00 & 0.00 & 0.00 \\ 
  AU RBA Cash Rate Target & 1.00 & 1.00 & 0.99 & 0.72 & 0.04 & 0.01 \\ 
  AU Wage Price Index QoQ & 0.03 & 0.00 & 0.00 & 0.00 & 0.00 & 0.00 \\ 
  AU Construction Work Done & 0.00 & 0.00 & 0.00 & 0.00 & 0.00 & 0.00 \\ 
  AU Private Capital Expenditure & 0.07 & 0.00 & 0.00 & 0.00 & 0.00 & 0.00 \\ 
  AU Inventories SA QoQ & 0.00 & 0.00 & 0.00 & 0.01 & 0.00 & 0.00 \\ 
  AU BoP Current Account Balance & 0.00 & 0.00 & 0.00 & 0.00 & 0.00 & 0.00 \\ 
  AU GDP SA QoQ & 1.00 & 0.00 & 0.00 & 0.00 & 0.00 & 0.00 \\ 
  AU Manpower Survey & 0.00 & 0.00 & 0.00 & 0.00 & 0.00 & 0.00 \\ 
  AU House Price Index QoQ & 0.00 & 0.00 & 0.00 & 0.01 & 0.00 & 0.00 \\ 
  AU RBA Statement on Monetary Policy & 0.00 & 0.00 & 0.01 & 0.00 & 0.00 & 0.00 \\ 
  AU S\&P Global Australia PMI Mfg & 0.00 & 0.00 & 0.00 & 0.00 & 0.00 & 0.00 \\ 
  AU S\&P Global Australia PMI Composite & 0.02 & 0.00 & 0.00 & 0.00 & 0.00 & 0.00 \\ 
  AU Home Loans Values MoM & 0.00 & 0.00 & 0.00 & 0.00 & 0.00 & 0.00 \\ 
  AU Private Sector Houses MoM & 0.11 & 0.00 & 0.00 & 0.00 & 0.00 & 0.00 \\ 
  AU Exports MoM & 0.08 & 0.00 & 0.00 & 0.00 & 0.00 & 0.00 \\ 
  AU RBA 3-Yr Yield Target & 0.00 & 0.00 & 0.00 & 0.00 & 0.00 & 0.00 \\ 
  AU CBA Household Spending YoY & 0.01 & 0.00 & 0.00 & 0.01 & 0.01 & 0.01 \\ 
  US S\&P Global US Manufacturing PMI & 0.00 & 0.00 & 0.00 & 0.00 & 0.06 & 0.02 \\ 
  US ISM Manufacturing & 0.24 & 0.10 & 0.04 & 0.00 & 0.00 & 0.00 \\ 
  US Construction Spending MoM & 0.26 & 0.13 & 0.02 & 0.00 & 0.00 & 0.00 \\ 
  US MBA Mortgage Applications & 0.00 & 0.00 & 0.00 & 0.00 & 0.00 & 0.00 \\ 
  US FOMC Meeting Minutes & 0.99 & 0.77 & 0.21 & 0.57 & 0.04 & 0.02 \\ 
  US Challenger Job Cuts YoY & 0.00 & 0.00 & 0.00 & 0.00 & 0.00 & 0.00 \\ 
  US ADP Employment Change & 0.29 & 0.01 & 0.00 & 0.00 & 0.00 & 0.00 \\ 
  US Initial Jobless Claims & 0.40 & 0.10 & 0.00 & 0.00 & 0.00 & 0.00 \\ 
  US S\&P Global US Services PMI & 0.01 & 0.00 & 0.00 & 0.01 & 0.00 & 0.01 \\ 
  US S\&P Global US Composite PMI & 0.00 & 0.00 & 0.00 & 0.01 & 0.00 & 0.01 \\ 
  US Langer Consumer Comfort & 0.00 & 0.00 & 0.00 & 0.00 & 0.00 & 0.00 \\ 
  US ISM Services Index & 0.17 & 0.00 & 0.00 & 0.00 & 0.00 & 0.00 \\ 
  US Trade Balance & 0.00 & 0.00 & 0.00 & 0.00 & 0.00 & 0.00 \\ 
  US Two-Month Payroll Net Revision & 0.02 & 0.01 & 0.20 & 0.00 & 0.06 & 0.51 \\ 
  US Change in Nonfarm Payrolls & 1.00 & 1.00 & 0.81 & 1.00 & 0.06 & 0.37 \\ 
  US Factory Orders & 0.00 & 0.00 & 0.00 & 0.00 & 0.00 & 0.00 \\ 
  US Durable Goods Orders & 0.00 & 0.00 & 0.00 & 0.00 & 0.00 & 0.00 \\ 
  US Consumer Credit & 0.00 & 0.00 & 0.00 & 0.00 & 0.00 & 0.00 \\ 
  US NFIB Small Business Optimism & 0.00 & 0.00 & 0.00 & 0.00 & 0.00 & 0.00 \\ 
  US Wholesale Inventories MoM & 0.00 & 0.00 & 0.00 & 0.00 & 0.00 & 0.00 \\ 
  US Wholesale Trade Sales MoM & 0.00 & 0.00 & 0.00 & 0.00 & 0.00 & 0.00 \\ 
  US JOLTS Job Openings & 0.00 & 0.00 & 0.00 & 0.00 & 0.00 & 0.00 \\ 
  US Import Price Index MoM & 0.00 & 0.00 & 0.00 & 0.00 & 0.00 & 0.00 \\ 
  US Monthly Budget Statement & 0.00 & 0.00 & 0.00 & 0.00 & 0.00 & 0.00 \\ 
  US PPI Final Demand MoM & 0.01 & 0.01 & 0.00 & 0.00 & 0.00 & 0.00 \\ 
  US Retail Sales Advance MoM & 0.99 & 0.02 & 0.00 & 0.01 & 0.00 & 0.00 \\ 
  US Business Inventories & 0.00 & 0.00 & 0.00 & 0.00 & 0.00 & 0.00 \\ 
  US U. of Mich. Sentiment & 0.00 & 0.00 & 0.00 & 0.02 & 0.01 & 0.00 \\ 
  US Empire Manufacturing & 0.00 & 0.00 & 0.00 & 0.00 & 0.00 & 0.00 \\ 
  US CPI MoM & 1.00 & 0.64 & 0.93 & 0.01 & 0.00 & 0.01 \\ 
  US Industrial Production MoM & 0.00 & 0.00 & 0.00 & 0.00 & 0.00 & 0.00 \\ 
  US NAHB Housing Market Index & 0.00 & 0.00 & 0.00 & 0.00 & 0.00 & 0.00 \\ 
  US Total Net TIC Flows & 0.00 & 0.00 & 0.00 & 0.00 & 0.00 & 0.00 \\ 
  US Housing Starts & 0.00 & 0.00 & 0.00 & 0.00 & 0.00 & 0.00 \\ 
  US Philadelphia Fed Business Outlook & 0.00 & 0.01 & 0.00 & 0.00 & 0.00 & 0.00 \\ 
  US Langer Economic Expectations & 0.00 & 0.00 & 0.00 & 0.00 & 0.00 & 0.00 \\ 
  US Existing Home Sales & 0.00 & 0.00 & 0.00 & 0.00 & 0.00 & 0.00 \\ 
  US Richmond Fed Manufact. Index & 0.00 & 0.00 & 0.00 & 0.00 & 0.00 & 0.00 \\ 
  US FHFA House Price Index MoM & 0.00 & 0.00 & 0.00 & 0.00 & 0.00 & 0.00 \\ 
  US Advance Goods Trade Balance & 0.00 & 0.00 & 0.00 & 0.00 & 0.00 & 0.00 \\ 
  US Chicago Fed Nat Activity Index & 0.00 & 0.00 & 0.00 & 0.00 & 0.00 & 0.00 \\ 
  US New Home Sales & 0.00 & 0.00 & 0.00 & 0.00 & 0.00 & 0.00 \\ 
  US Leading Index & 0.00 & 0.00 & 0.00 & 0.00 & 0.00 & 0.00 \\ 
  US Kansas City Fed Manf. Activity & 0.00 & 0.00 & 0.00 & 0.00 & 0.00 & 0.00 \\ 
  US GDP Annualized QoQ & 0.02 & 0.00 & 0.00 & 0.00 & 0.00 & 0.00 \\ 
  US Core PCE Price Index MoM & 0.00 & 0.00 & 0.00 & 0.00 & 0.00 & 0.00 \\ 
  US Pending Home Sales MoM & 0.00 & 0.00 & 0.00 & 0.00 & 0.00 & 0.00 \\ 
  US Dallas Fed Manf. Activity & 0.00 & 0.00 & 0.00 & 0.00 & 0.00 & 0.01 \\ 
  US Employment Cost Index & 0.01 & 0.01 & 0.00 & 0.00 & 0.00 & 0.00 \\ 
  US S\&P CoreLogic CS 20-City MoM SA & 0.00 & 0.00 & 0.00 & 0.00 & 0.00 & 0.00 \\ 
  US S\&P CoreLogic CS 20-City NSA Index & 0.00 & 0.00 & 0.00 & 0.00 & 0.00 & 0.00 \\ 
  US MNI Chicago PMI & 0.00 & 0.00 & 0.00 & 0.00 & 0.00 & 0.00 \\ 
  US Conf. Board Consumer Confidence & 0.00 & 0.00 & 0.00 & 0.00 & 0.00 & 0.00 \\ 
  US FOMC Rate Decision (Upper Bound) & 1.00 & 0.98 & 1.00 & 0.88 & 0.79 & 0.06 \\ 
  US Nonfarm Productivity & 0.00 & 0.00 & 0.00 & 0.00 & 0.00 & 0.00 \\ 
  US Mortgage Delinquencies & 0.00 & 0.00 & 0.00 & 0.00 & 0.00 & 0.00 \\ 
  US House Price Purchase Index QoQ & 0.00 & 0.00 & 0.00 & 0.00 & 0.00 & 0.00 \\ 
  US U.S. Federal Reserve Releases Beige Book & 0.00 & 0.00 & 0.00 & 0.00 & 0.00 & 0.00 \\ 
  US Household Change in Net Worth & 0.00 & 0.00 & 0.00 & 0.00 & 0.00 & 0.00 \\ 
  US Current Account Balance & 0.00 & 0.00 & 0.01 & 0.00 & 0.00 & 0.00 \\ 
  US Export Price Index MoM & 0.00 & 0.00 & 0.00 & 0.00 & 0.00 & 0.00 \\ 
  US Fed Interest on Reserve Balances Rate & 0.00 & 0.01 & 0.02 & 0.06 & 0.06 & 0.06 \\ 
  US ISM Services Employment & 0.02 & 0.02 & 0.03 & 0.03 & 0.02 & 0.03 \\ 
  US New York Fed Services Business Activity & 0.03 & 0.02 & 0.03 & 0.03 & 0.03 & 0.02 \\ 
  US Philadelphia Fed Non-Manufacturing Activity & 0.03 & 0.02 & 0.03 & 0.02 & 0.03 & 0.03 \\ 
  US Richmond Fed Business Conditions & 0.02 & 0.03 & 0.02 & 0.02 & 0.03 & 0.03 \\ 
  US Kansas City Fed Services Activity & 0.03 & 0.03 & 0.02 & 0.03 & 0.03 & 0.03 \\ 
  US Dallas Fed Services Activity & 0.03 & 0.03 & 0.03 & 0.02 & 0.03 & 0.02 \\ 
  US NY Fed 1-Yr Inflation Expectations & 0.02 & 0.02 & 0.03 & 0.02 & 0.02 & 0.02 \\ 
   \hline
   \caption{Posterior mean of $\pi$ for all events}
\end{longtable}